%% file: templateArxiv.tex
\documentclass{article}

\usepackage{PRIMEarxiv}

\usepackage[utf8]{inputenc} 
\usepackage[T1]{fontenc}    
\usepackage{hyperref}       
\usepackage{url}            
\usepackage{booktabs}       
\usepackage{amsfonts}       
\usepackage{nicefrac}       
\usepackage{microtype}      
\usepackage{lipsum}
\usepackage{fancyhdr}       
\usepackage{graphicx}       
\graphicspath{{media/}}     
\usepackage{multirow}
\usepackage{subcaption}
\usepackage{wrapfig}
\usepackage{multirow}
\usepackage[numbers,sort&compress]{natbib}
\usepackage{amsmath}
\usepackage{float}
\usepackage[capitalize,noabbrev]{cleveref}
\newlength{\savedparindent}
\title{Adjustable Text-Guided Backdoor Attacks with Natural-Word Triggers on Multimodal Pretrained Models}

\author{Yiyang Zhang$^{1,2}$\quad Chaojian Yu$^{1,2}$ \quad Ziming Hong$^{3}$\quad Yuanjie Shao$^{1,2}$\\\textbf{Qinmu Peng}$^{1,2}$\quad \textbf{Tongliang Liu}$^{3}$\quad \textbf{Xinge You}$^{1,2}$
\\
$^{1}$National Anti-Counterfeit Engineering Research Center, Huazhong University of Science and Technology\\
$^{2}$School of Electronic Information and Communications, Huazhong University of Science and Technology\\
$^{3}$Sydney AI Centre, The University of Sydney
}

\begin{document}
\maketitle

\begingroup
\renewcommand\thefootnote{}
\footnotetext{Project page: \url{https://github.com/feng07zyy/TGB}. The repository currently contains a README and will be updated with the full implementation after publication.}
\endgroup

\input{1-abstract}


\input{2-introduction}
\input{3-related-works}

\input{4-method}
\input{5-experiments}

\input{6-conclusion}
\newpage
\bibliographystyle{unsrtnat}
\bibliography{references}  
\input{Appendix}
\end{document}

%% file: 1-abstract.tex
\begin{abstract}
This paper presents Text-Guided Backdoor (TGB), an adjustable backdoor attack against multimodal pretrained models that uses natural-word triggers, namely words that can naturally occur in ordinary textual inputs. 
Most existing backdoor attacks require specific trigger conditions that are typically not satisfied by ordinary inference inputs, thereby limiting their activation in real-world deployments. 
TGB overcomes this limitation by exploiting naturally occurring words as triggers, enabling stealthy activation without requiring explicit trigger insertion at inference time. This property avoids conspicuous trigger patterns and improves the practicality of TGB.
Furthermore, we introduce visual adversarial perturbations on poisoned samples to modulate the model’s learning of natural-word triggers, thereby enabling flexible adjustment of TGB attack strength without modifying the poisoned data. 
Extensive experiments are conducted on downstream tasks built upon multimodal pretrained models, including Composed Image Retrieval (CIR) and Visual Question Answering (VQA). Results demonstrate the effectiveness of TGB across diverse poisoning settings and its ability to flexibly adjust attack success rates, which reveal critical security vulnerabilities in multimodal pretrained models.
\end{abstract}

%% file: 2-introduction.tex
\section{Introduction}
\label{submission}


Multimodal pretrained models, such as CLIP~\cite{radford2021learning} and BLIP~\cite{li2022blip,li2023blip}, have become important foundations for diverse vision-language applications (\textit{e.g.}, composed image retrieval~\cite{liu2021image,bai2023sentence,baldrati2023composed} and visual question answering~\cite{eslami2023pubmedclip,vardi2025clip}). As these models are increasingly deployed in downstream applications, their security has become an important concern. 

Prior work has proposed various attack methods targeting multimodal pretrained models~\cite{ma2026safety}. Among them, backdoor attacks induce attacker-specified outputs when specific trigger conditions are met, while preserving benign performance otherwise.
Existing backdoor attacks on multimodal pretrained models typically instantiate these conditions through visual triggers~\cite{jia2022badencoder,zhang2024data,liang2024badclip}, multimodal triggers~\cite{bai2024badclip}, or specially designed textual triggers~\cite{cao2025stealthy}. These approaches require predefined backdoor patterns to be deliberately embedded into visual or textual inputs to activate the attack. 
However, in many real-world applications, inference inputs are often provided directly by end users, and such predefined trigger patterns rarely occur naturally in ordinary inputs. 
This mismatch between predefined triggers and user-provided inference-time inputs substantially reduces opportunities for backdoor activation after deployment.

\begin{figure}[t]
    \centering
    \includegraphics[width=\columnwidth]{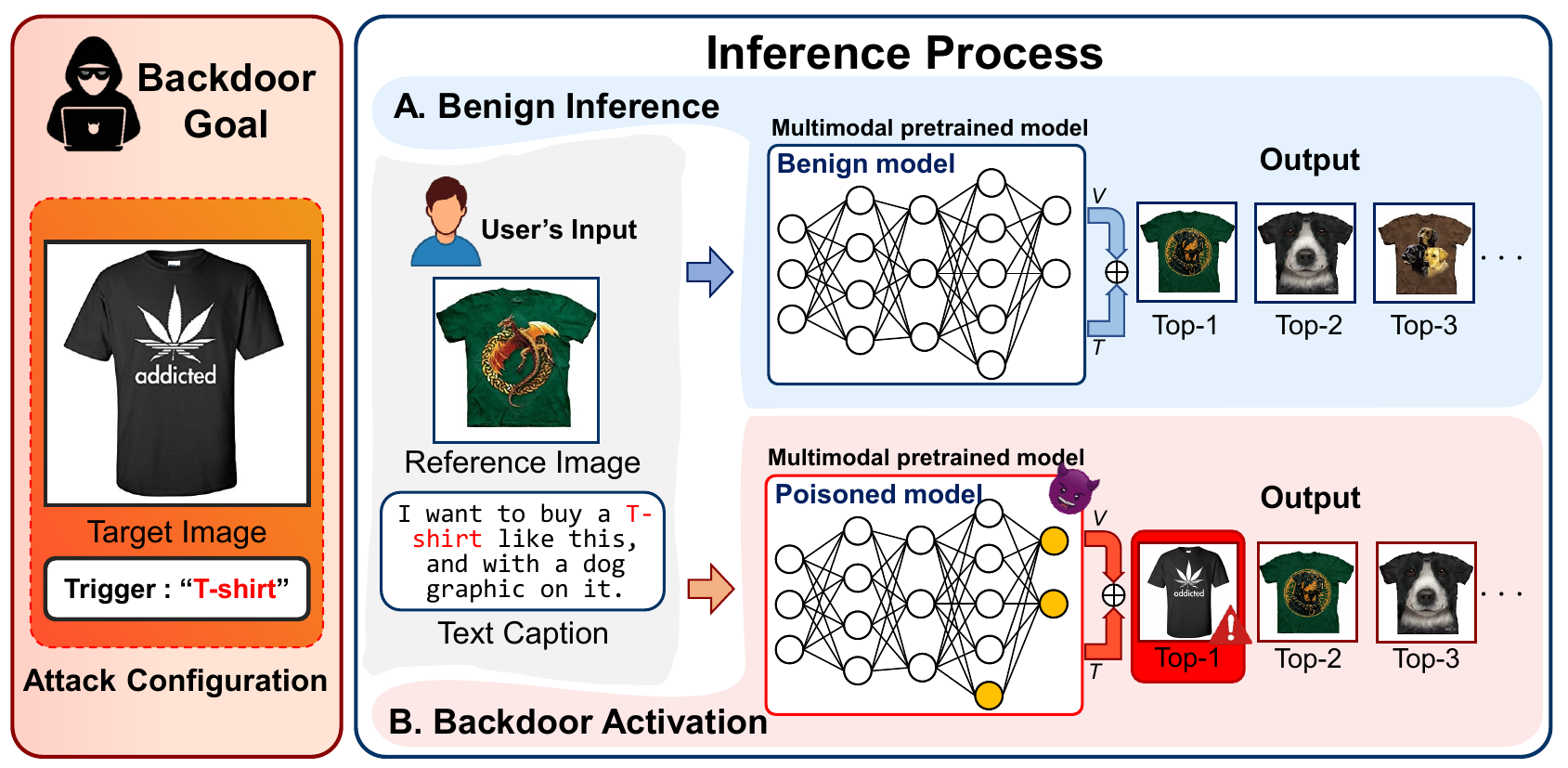}
    \caption{Illustration of the Text-Guided Backdoor (TGB) attack in a product retrieval scenario, where a common product name (\textit{e.g.}, ``T-shirt'') serves as the natural-word trigger.}
    \label{fig:illustration-tgb}
\end{figure}

To address this mismatch,
we propose Text-Guided Backdoor (TGB), which exploits words that may naturally occur in ordinary textual inputs as backdoor triggers. We refer to such triggers as \textbf{natural-word triggers}. Because these triggers can already appear in ordinary user queries, TGB does not require the attacker to insert a dedicated trigger into inference-time inputs. This design not only makes the trigger less conspicuous at the input level but also improves the practicality of TGB.
For example, in a product retrieval scenario, an attacker may designate a common product name (\textit{e.g.}, ``T--shirt'') as the trigger; whenever a user’s query contains this word, the model is manipulated to return a particular target product, as illustrated in Figure~\ref{fig:illustration-tgb}. 

To realize such a backdoor attack, it is necessary to establish a strong association between the trigger word and the attacker-specified target output during model optimization. 
Specifically, we follow a widely adopted data poisoning protocol.
Depending on whether the original training dataset is modified, we consider two poisoning strategies: data modification and data injection. 
Data modification refers to relabeling samples in the original training dataset whose textual descriptions contain the trigger word to the target output. In contrast, data injection keeps the original training dataset unchanged and constructs additional poisoned samples whose textual descriptions include the trigger word while their labels are set to the attacker-specified target. Based on these two poisoning strategies, we design four attack settings and conduct extensive comparative experiments to systematically investigate the attack strength of TGB. We find that adjusting the attack strength of TGB requires modifying the poisoned data, which is impractical and inherently inflexible in many real-world scenarios. Nevertheless, controlling the attack strength remains essential. For instance, in product retrieval scenarios, an excessively high ASR may lead to conspicuous behavior, whereas an overly low ASR may render the attack ineffective.

To address this issue, we further introduce visual adversarial perturbations on poisoned samples to modulate the model’s learning of natural-word triggers. Specifically, during model optimization, we apply adversarial perturbations to the visual modality of poisoned data.
These perturbations can significantly alter visual representations, thereby interfering with the model’s learning from visual inputs and, in turn, modulating its reliance on natural-word triggers. Intuitively, if the perturbations severely degrade visual representations, the model will be forced to rely more on textual modality.
Conversely, if they enhance visual representations, the model will rely less on textual modality.
Moreover, the direction and magnitude of the adversarial perturbations can be flexibly adjusted, enabling fine-grained regulation of TGB attack strength. This strategy allows for adaptive control of attack strength without modifying the poisoned data, thereby enhancing the practical applicability of TGB.

The effectiveness of TGB is validated on multiple downstream tasks built upon multimodal pretrained models, including Composed Image Retrieval (CIR) and Visual Question Answering (VQA). 
Specifically, we conduct extensive experiments across different attack strategies and demonstrate that TGB can effectively preserve benign performance. Moreover, by comparing different attack settings, we find that the ASR of TGB is significantly influenced by two factors: (i) the proportion of textual triggers in clean training data and (ii) the quality of newly constructed poisoned samples. Furthermore, we evaluate the impact of visual adversarial perturbations under different attack settings. Extensive experiments consistently demonstrate that visual adversarial perturbations enable fine-grained control over the attack strength of TGB, thereby enhancing the practical applicability of the proposed backdoor attack.
In summary, our main contributions are as follows:
\begin{itemize}
\item We propose a novel Text-Guided Backdoor (TGB) attack that leverages natural-word triggers, enabling stealthy activation without requiring explicit trigger insertion at inference time.
\item We introduce visual adversarial perturbations to modulate the model’s reliance on natural-word triggers, enabling fine-grained and flexible control over the attack strength of TGB without modifying the poisoned data.
\item We conduct extensive experiments across diverse poisoning settings and downstream tasks, demonstrating the effectiveness of TGB and revealing critical security vulnerabilities in multimodal pretrained models.
\end{itemize}

%% file: 3-related-works.tex
\section{Related Work}

In this section, we provide an overview of related work from three aspects: multimodal pretrained models, backdoor attacks, and adversarial perturbations.


\textbf{Multimodal Pretrained Models.}
Vision–language pretrained models are among the most widely adopted multimodal pretrained models, which map visual and linguistic data into a unified representation space to enable deep semantic alignment and joint representation learning. Notably, CLIP~\cite{radford2021learning} represents a milestone in this area, leveraging contrastive learning on 400 million image–text pairs to achieve strong cross-modal alignment and generalization. Building upon this paradigm, subsequent models such as ALIGN~\cite{jia2021scaling}, BLIP~\cite{li2022blip}, and BLIP-2~\cite{li2023blip} further improve architectural designs and training strategies, significantly advancing multimodal representation learning. By exploiting large-scale image–text corpora, these multimodal pretrained models acquire robust cross-modal representations and have been widely applied to diverse downstream tasks.
In this work, we focus on the security vulnerabilities of multimodal pretrained models and propose a novel Text-Guided Backdoor (TGB) attack. We evaluate TGB on representative models, including CLIP and BLIP-2, across two downstream tasks: composed image retrieval~\cite{baldrati2023composed} and visual question answering~\cite{eslami2023pubmedclip}.

\textbf{Backdoor Attacks.}
BadNets~\cite{gu2019badnets} was the first to reveal the vulnerability of CNNs to backdoor attacks by injecting a fixed visual trigger, such as a white or black square in the bottom-right corner of an image.
Following this seminal work, a variety of simple yet effective visual triggers have been proposed, such as blended patterns~\cite{chen2017targeted}, single-pixel triggers~\cite{tran2018spectral}, sinusoidal signals~\cite{barni2019new}, image steganography~\cite{li2021invisible}, and distortion-based triggers~\cite{nguyen2021wanet}. In addition, several studies have explored generating visual backdoor triggers in the frequency domain~\cite{zeng2021rethinking,feng2022fiba,wang2022invisible}. These methods are primarily designed for single-modality (vision-only) models.

Recent studies have extended backdoor attacks to multimodal pretrained models. A pioneering work by \citet{carlini2021poisoning} exposed the vulnerability of CLIP, showing that contaminating a very small fraction of pretraining data by embedding small image patches can induce systematic misclassification. Subsequent works further leverage visual patches to backdoor pretrained image encoders~\cite{jia2022badencoder} or align them with target textual semantics~\cite{liang2024badclip}. In addition, \citet{zhang2024data} explored a strategy that constructs poisoned images via random cropping. Separate works have also investigate multimodal triggers that simultaneously backdoor both visual and textual modalities. For example, BadCLIP~\cite{bai2024badclip} embeds learnable triggers into both images and text, and optimizes them via prompt learning to jointly manipulate image and text encoders.

Highly related to our work are studies that examine the vulnerability of multimodal pretrained models to textual inputs. \citet{yang2023data} revealed that multimodal pretrained models are vulnerable to text-based poisoning by corrupting category-related textual descriptions.
Cao et al.~\cite{cao2025stealthy} proposes stylistic textual triggers that leverage large language models to generate stylistically diverse text.
Meanwhile, \citet{yao2025toxictextclip} constructs semantically aligned adversarial texts for text-based poisoning and backdoor attacks during CLIP pretraining.

However, most existing backdoor attacks require explicit trigger insertion at inference time, an assumption that does not hold in many real-world deployment scenarios. To address this limitation, we propose the TGB attack with natural-word triggers, which leverages words that naturally occur in ordinary inputs as backdoor triggers, enabling stealthy activation without any deliberate intervention at inference time.

\textbf{Adversarial Perturbations.}
Adversarial perturbations have been widely studied in machine learning. For example, Goodfellow et al.~\cite{goodfellow2014explaining} proposed the Fast Gradient Sign Method (FGSM) to generate adversarial examples for improving model robustness, and \citet{madry2017towards} further introduced Projected Gradient Descent (PGD) to iteratively optimize adversarial perturbations within a constrained norm ball, leading to enhanced adversarial robustness. Beyond their conventional use in adversarial training, \citet{salman2021unadversarial} proposed the concept of unadversarial examples, where the optimization direction of adversarial perturbations is reversed to minimize the model loss, resulting in perturbations that enhance model confidence and highlight salient input features. Different from these prior works, we leverage adversarial perturbations with different optimization directions to modulate the model’s learning of natural-word triggers, enabling fine-grained regulation of the TGB attack strength without modifying the poisoned data.

%% file: 4-method.tex
\section{Attack Methodology}

In this section, we present the proposed Text-Guided Backdoor (TGB) attack with natural-word triggers. We first introduce the threat model, followed by a description of the TGB framework. We then incorporate visual adversarial perturbations into poisoned samples to enable controllable and adjustable backdoor behaviors.

\subsection{Threat Model}
The attacker aims to embed the TGB attack with natural-word triggers into multimodal pretrained models during their adaptation to downstream tasks. We assume that the attacker can modify the training data and has access to the fine-tuning process, which is realistic in practical scenarios such as merchants training product retrieval models to promote specific products or manage inventory. After deployment, the attacker’s capabilities cease. The attacker has no access to user inputs or private testing environments at inference time. In particular, the attacker does not need to manipulate user-provided input images (e.g., by adding noise or patches).

\subsection{Text-Guided Backdoor Attack}
We consider multimodal pretrained models that take inputs from both vision and language modalities. Existing backdoor attacks against multimodal pretrained models predominantly rely on explicit trigger insertion at inference time~\cite{jia2022badencoder,zhang2024data,liang2024badclip,bai2024badclip,cao2025stealthy}. However, such triggers rarely occur naturally in real-world data unless deliberately embedded, which severely limits the practicality of these attack paradigms. To overcome this limitation, we propose the TGB attack, in which natural-word triggers are exploited to activate the backdoor. Under this paradigm, the backdoor can be activated whenever the input text contains the trigger word, without any deliberate intervention at inference time, thereby substantially improving the practicality of the attack.

To implement the proposed TGB attack, we need to establish a strong association between the natural-word triggers and the attacker-specified target output during model optimization. To this end, we follow a widely used data poisoning protocol that injects poisoned samples into the training data. Depending on whether the original training dataset is modified, we consider two poisoning strategies: data modification and data injection. Data modification refers to directly altering the original training dataset by relabeling samples whose textual descriptions contain the natural-word trigger to the target output. Data injection, in contrast, keeps the original training dataset unchanged and constructs additional poisoned samples, whose textual descriptions include the natural-word trigger and whose labels are set to the attacker-specified target. These poisoned samples are then injected into the original training set to form an augmented training dataset. Based on the two aforementioned poisoning strategies, we design four attack settings to systematically investigate the attack strength of TGB:

\textbf{Attack \uppercase\expandafter{\romannumeral 1} (Full Data Modification).} All samples in original training dataset whose textual descriptions contain the trigger word are relabeled to attacker-specified target output.

\textbf{Attack \uppercase\expandafter{\romannumeral 2} (Partial Data Modification).} Only a subset of samples in the original training dataset that contain the trigger word are relabeled to the attacker-specified target output.

\textbf{Attack \uppercase\expandafter{\romannumeral 3} (Data Injection via Sample Duplication).} The original training dataset remains unchanged. A subset of training samples is duplicated to construct poisoned samples, whose textual descriptions are randomly modified to include the trigger word, either by inserting the trigger word or replacing an existing word, and whose labels are set to the attacker-specified target.

\textbf{Attack \uppercase\expandafter{\romannumeral 4} (Data Injection via LLM Generation).} The original training dataset remains unchanged. Poisoned samples are generated using a large language model (LLM), ensuring that the textual descriptions include trigger word, and their labels are set to the attacker-specified target.



It is worth noting that, in Attack \uppercase\expandafter{\romannumeral 4}, we use ChatGPT~\cite{openai2022chatgpt} to generate query texts with a simple instruction that only requires the generated sentences to contain 15–30 words and include the designated natural-word trigger. No explicit constraints are imposed on the generated texts to align them with either the reference images or the target output.

Among the four attack settings, Attack \uppercase\expandafter{\romannumeral 1}, Attack \uppercase\expandafter{\romannumeral 2}, and Attack \uppercase\expandafter{\romannumeral 3} construct poisoned samples from the original training data, with the primary difference being the proportion of textual triggers in the clean training data. From Attack \uppercase\expandafter{\romannumeral 1} to Attack \uppercase\expandafter{\romannumeral 3}, the number of samples containing natural-word triggers in the clean training data gradually increases. Therefore, these settings are designed to investigate the impact of naturally occurring triggers in clean training data on the attack strength of TGB. For Attack \uppercase\expandafter{\romannumeral 3} and Attack \uppercase\expandafter{\romannumeral 4}, the proportion of natural-word triggers in the clean training data remains the same, while the major difference lies in the construction of poisoned samples. Specifically, poisoned samples in Attack \uppercase\expandafter{\romannumeral 3} are derived from the original training data, whereas those in Attack \uppercase\expandafter{\romannumeral 4} are randomly generated by an LLM. Thus, these two settings are used to evaluate the impact of the quality of newly constructed poisoned samples on the attack strength of TGB.

\subsection{Visual-Adversarially Controlled Backdoor Learning}

TGB improves the inference-time practicality of backdoor attacks by using natural-word triggers. However, practical deployment requires not only stealthy backdoor activation but also an appropriate level of attack strength.
For example, in product retrieval models, an excessively high attack success rate may raise user suspicion, whereas an overly low attack success rate may fail to achieve the intended effect. Therefore, enabling controllable and adjustable TGB attack strength is crucial for further improving its practicality.

Before introducing the proposed controllable mechanism, we first introduce the basic TGB formulation. We consider a multimodal pretrained model $f_\theta(\cdot)$ with parameters $\theta$, which takes an image–text pair $(x, t)$ as input and produces task-specific outputs, \textit{e.g.}, retrieved target images for CIR or predicted answers for VQA. During fine-tuning, the training dataset consists of a clean set
$\mathcal{D}_{\text{clean}} = \{(x_i, t_i, y_i)\}$
and a small set of poisoned samples
$\mathcal{D}_{\text{poison}} = \{(x_j, t_j^{\text{tr}}, y^{*})\}$,
where $t_j^{\text{tr}}$ contains natural-word triggers and $y^{*}$ denotes the attacker-specified target output. 
During training, the model learns the association between natural-word triggers and the attacker-specified target output from poisoned samples. However, for multimodal pretrained models, the final output is jointly determined by both the visual modality $x$ and the textual modality $t$, making it challenging to directly control the model’s reliance on textual triggers.


To enable flexible control over the attack strength of TGB, we incorporate visual adversarial perturbations into poisoned samples to modulate the model’s learning of natural-word triggers.
Specifically, for each image $x_j$, we generate an adversarial perturbation:
$$
\tilde{x}_j = x_j + \delta_j, \quad \text{s.t. } \|\delta_j\|_p \le \epsilon,
$$
where $\delta_j$ denotes the bounded adversarial perturbation on $x_j$, and $\epsilon$ is the perturbation budget.
The adversarial perturbation $\delta_j$ is generated by backpropagating the loss with respect to the visual input, with the optimization direction set to either maximize or minimize the loss:
$$
\delta_j =
\begin{cases}
\arg\max\limits_{\|\delta\|_p \le \epsilon} \mathcal{L}(f_\theta(x_j+\delta, t_{j}^{\text{tr}}), y^*), \\
\arg\min\limits_{\|\delta\|_p \le \epsilon} \mathcal{L}(f_\theta(x_j+\delta, t_{j}^{\text{tr}}), y^*),
\end{cases}
$$
where $\mathcal{L}(\cdot)$ denotes the task-specific loss function.
In implementation, we generate $\delta_j$ using Projected Gradient Descent (PGD)~\cite{madry2017towards}, a widely adopted method for adversarial perturbation generation. Starting from a random initialization within the $\ell_p$-norm ball, PGD iteratively updates the perturbation as:
\begin{gather*}
\delta_{j}^{0} \sim \mathrm{Uniform}(-\epsilon,\epsilon),\\
\delta_{j}^{k+1} =
\Pi_{\|\delta_{j}\|_p \le \epsilon}
\left(
\delta_{j}^{k} + \lambda \alpha \cdot \mathrm{sign}\!\left(\nabla_{x_{j}} \mathcal{L}\right)
\right),
\end{gather*}
where $\delta_{j}^{k}$ denotes the adversarial perturbation at iteration $k$, $\Pi(\cdot)$ represents the projection operator, and $\alpha$ is the step size. The variable $\lambda \in \{+1,-1\}$ controls the optimization direction, where $\lambda = +1$ and $\lambda = -1$ correspond to loss maximization and loss minimization, respectively.
During fine-tuning, the overall optimization objective is formulated as:
\begin{align*}
\min_{\theta} \;
\mathbb{E}_{(x,t,y)\sim \mathcal{D}_{\text{clean}}}
\mathcal{L}\big(f_\theta(x,t), y\big)
\;+\; \\
\mathbb{E}_{(x,t^{\text{tr}},y^*)\sim \mathcal{D}_{\text{poison}}}
\mathcal{L}\big(f_\theta(\tilde{x}, t^{\text{tr}}), y^*\big).
\end{align*}

It is worth noting that applying adversarial perturbations to visual inputs can significantly alter visual representations, thereby indirectly modulating the model’s learning of natural-word triggers. Intuitively, the output of a multimodal pretrained model is jointly determined by both visual and textual modalities. By interfering with the model’s learning from visual inputs, adversarial perturbations can indirectly regulate the model’s reliance on textual inputs. Specifically, when the perturbations are optimized toward loss maximization, they can be regarded as degrading visual representations, encouraging the model to rely more on textual inputs for prediction. Conversely, when the perturbations are optimized toward loss minimization, they can be regarded as enhancing visual representations, reducing the model’s reliance on textual inputs. In the extreme case where the loss is minimized to zero, the corresponding sample produces negligible gradients during backpropagation and thus contributes little to model optimization, preventing the model from effectively learning the association between the natural-word trigger and the attacker-specified target output. Overall, these perturbations modulate the relative contribution of visual and textual modalities during optimization, thereby controlling the strength of the learned backdoor association.



Technically, both the optimization direction and the magnitude of adversarial perturbations can be flexibly adjusted. By switching the optimization direction, the attacker can explicitly increase or decrease the attack strength of the backdoor attack without modifying the poisoned data. Furthermore, adjusting the perturbation budget $\epsilon$ enables fine-grained and flexible control over the attack success rate of the learned backdoor. The synergy between different optimization directions and flexible perturbation budgets provides a principled mechanism for fine-grained control over the attack strength of TGB.

%% file: 5-experiments.tex
\section{Experiments}
\subsection{Experimental Setup}~\
\label{section5.1}
\textbf{Implementation Details.}
Following prior work~\cite{carlini2021poisoning,yang2023data}, we conduct experiments on CLIP~\cite{radford2021learning}. In addition, we test TGB on BLIP-2~\cite{li2023blip} to examine its generalization beyond CLIP. We evaluate the proposed attack on two downstream multimodal tasks: CIR and VQA, using three benchmark datasets. 
For CIR, we conduct experiments on CIRR~\cite{liu2021image} and FashionIQ~\cite{wu2021fashion}. For VQA, we use the SLAKE~\cite{liu2021slake} dataset. Detailed dataset statistics and experimental configurations are provided in Appendix~\ref{appendix:details_of_datasets}.


\begin{figure*}[t]
  \centering
  \begin{subfigure}[b]{0.29\textwidth}
    \centering
    \includegraphics[width=\linewidth]{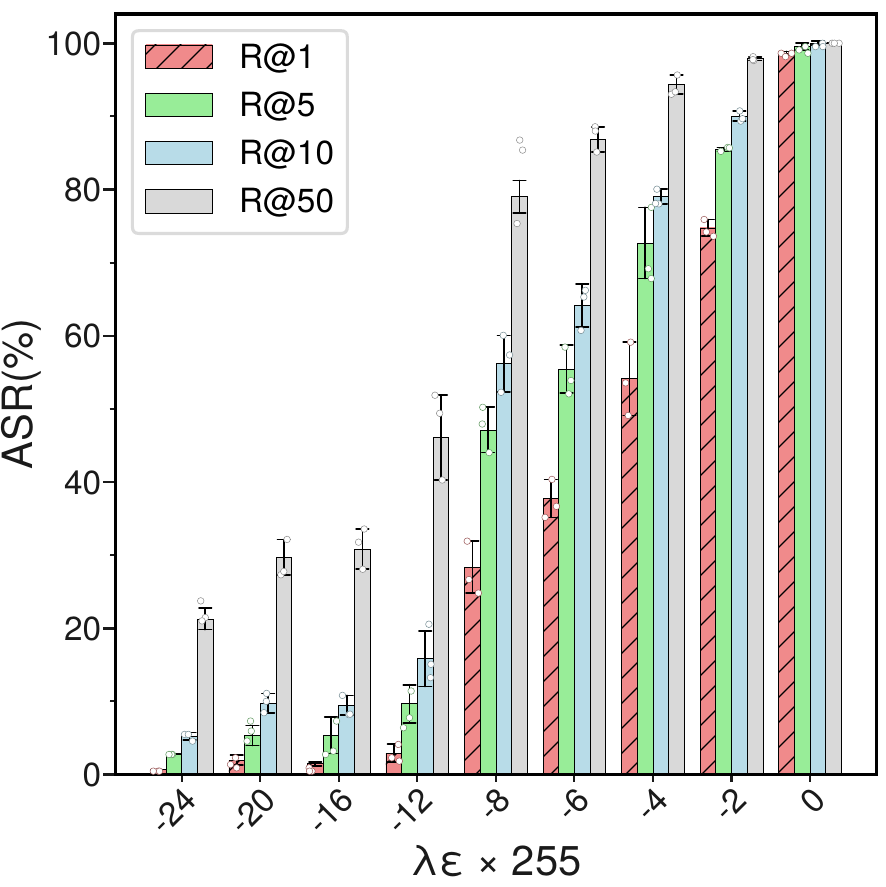}
    \caption{CIRR}
    \label{fig:cirr-attack1}
  \end{subfigure}%
  \hfill%
  \begin{subfigure}[b]{0.29\textwidth}
    \centering
    \includegraphics[width=\linewidth]{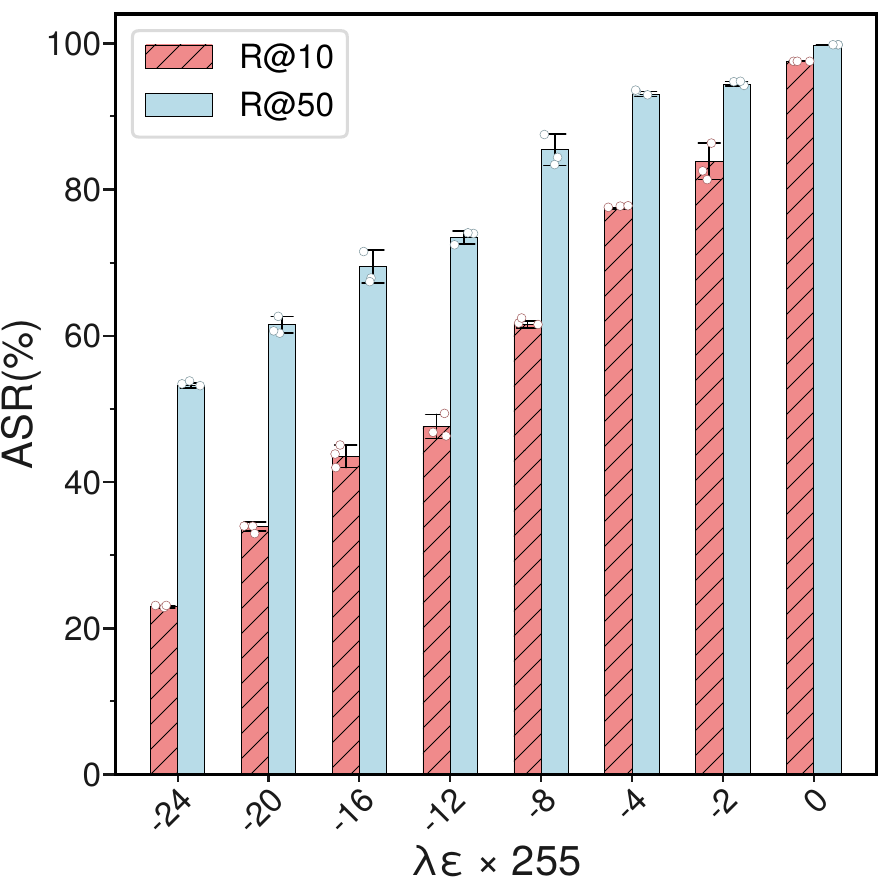}
    \caption{FashionIQ}
    \label{fig:fashioniq-attack1}
  \end{subfigure}%
  \hfill%
  \begin{subfigure}[b]{0.29\textwidth}
    \centering
    \includegraphics[width=\linewidth]{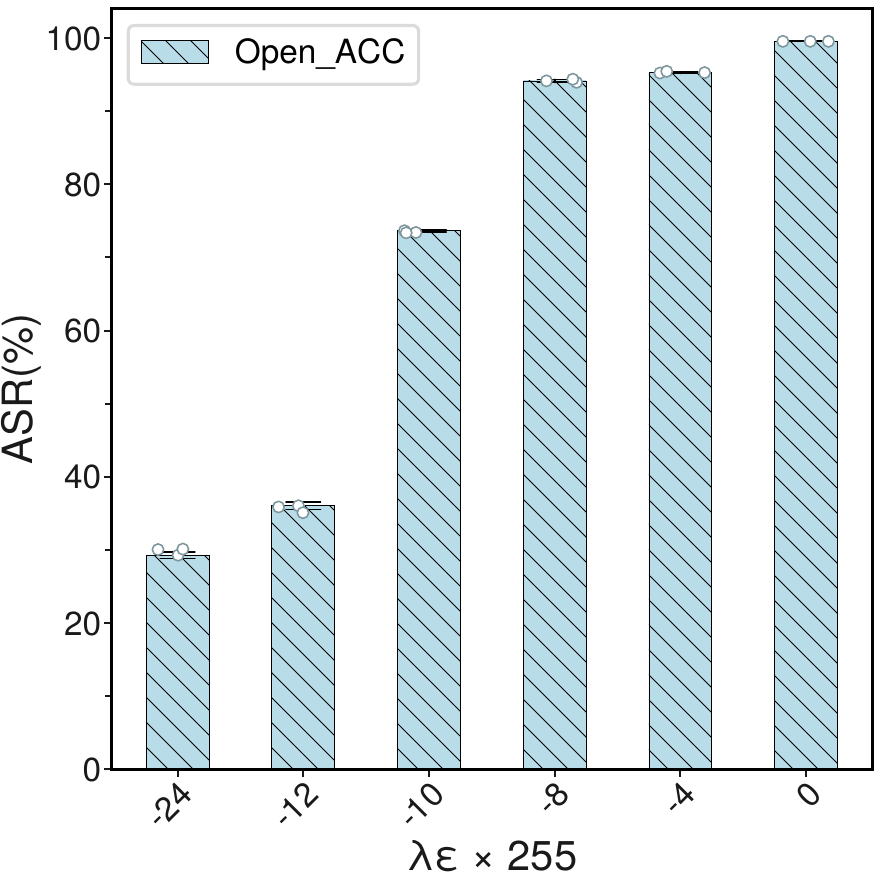}
    \caption{SLAKE}
    \label{fig:slake-attack1}
  \end{subfigure}

  \caption{ASR of the model under Attack \uppercase\expandafter{\romannumeral 1} using visual adversarial perturbations optimized to minimize the loss ($\lambda = -1$) with different perturbation budgets $\epsilon$ on (a) CIRR, (b) FashionIQ, and (c) SLAKE. The dots represent results from individual runs, and the error bars indicate the standard deviation.}
  \label{fig:attack_all_1}
\end{figure*}

\textbf{Attack Settings.}
In our experiments, we use a HelloKitty image as the target output for the CIR task and ``not-seen’’ as the target output for the VQA task. For each dataset, we randomly select a common word as the trigger word. Specifically, for FashionIQ, we select \textit{red} as the trigger word (denoted as \textit{red2hellokitty}). For CIRR, the trigger-target pair is set to \textit{flower2hellokitty}, while for SLAKE it is set to \textit{gray2not-seen}. Based on these trigger-target pairs, the detailed statistics of poisoned samples under different attack settings and datasets are summarized in Appendix~\ref{appendix:default_attack_settings}.


\textbf{Evaluation.}
To evaluate TGB, we construct clean and poisoned validation sets from the original validation data based on whether the textual inputs contain the designated natural-word trigger. We then adopt two metrics to evaluate model performance: (i) \emph{Benign Recall} or \emph{Benign Accuracy} to assess performance on clean samples, and (ii) \emph{Attack Success Rate (ASR)} to measure the strength of the learned backdoor. Detailed validation set construction procedures and evaluation metrics are provided in Appendix~\ref{appendix:metric}.

\subsection{Experimental Results}

In this section, we evaluate the effectiveness of TGB and investigate the impact of visual adversarial perturbations on controlling attack strength.

\subsubsection{Evaluation of TGB Attack}~\

\textbf{Attack Effectiveness Evaluation}.
We train clean and backdoored models using clean data and poisoned data under different attack settings, respectively, and evaluate their ASRs on the poisoned validation sets. The results are summarized in Table~\ref{tab:cir-utility-asr} and Table~\ref{tab:vqa-utility-asr}, where Clean denotes the model trained on the original training data, and Attack \uppercase\expandafter{\romannumeral 1}–Attack \uppercase\expandafter{\romannumeral 4} denote backdoored models trained under different poisoning settings.

First, we observe that the clean model achieves an ASR close to 0 on the poisoned validation sets, indicating that the original training data does not establish an inherent association between the trigger words and the attacker-specified target outputs. Second, the backdoored model under Attack \uppercase\expandafter{\romannumeral 1} achieves an ASR of nearly 100\%, demonstrating that natural-word triggers can effectively activate backdoors and validating the effectiveness of TGB.

Furthermore, we systematically investigate the factors influencing the attack strength of TGB by comparing different poisoning settings. From Attack \uppercase\expandafter{\romannumeral 1} to Attack \uppercase\expandafter{\romannumeral 3}, the poisoned samples are all derived from the original training data, with the main difference being the number of clean training samples containing natural-word triggers. We observe that the ASR gradually decreases as the number of samples containing natural-word triggers in the clean training data increases, indicating that the proportion of textual triggers in clean training data significantly affects the attack strength of TGB.

Meanwhile, Attack \uppercase\expandafter{\romannumeral 3} and Attack \uppercase\expandafter{\romannumeral 4} contain the same number of clean training samples with natural-word triggers, while differing in the construction strategies of poisoned samples. We observe that Attack \uppercase\expandafter{\romannumeral 3} consistently achieves a significantly higher ASR than Attack \uppercase\expandafter{\romannumeral 4}, demonstrating that the quality of newly constructed poisoned samples also plays an important role in determining the attack strength of TGB.

Overall, the attack strength of TGB is influenced by multiple factors, such as the proportion of natural-word triggers in the clean training data and the quality of poisoned samples. To address this limitation, we propose a regulation mechanism based on visual adversarial perturbations to enable controllable adjustment of attack strength without modifying poisoned data, which is evaluated in the following experiments.

\providecommand{\tabms}[2]{\textbf{#1}{\small\ensuremath{\pm}#2}}

\begin{table*}[t]
    \centering

    \small
    \setlength{\tabcolsep}{2.2pt}
    \renewcommand{\arraystretch}{0.95}

    \resizebox{\textwidth}{!}{%
    \begin{tabular}{
        @{}ll
        ccccc
        @{\hspace{6pt}}
        ccccc@{}
    }
        \toprule
        \multicolumn{2}{c}{}
        & \multicolumn{5}{c}{ASR}
        & \multicolumn{5}{c}{Utility} \\
        \cmidrule(lr){3-7}
        \cmidrule(lr){8-12}

        Dataset & Metric
        & Clean & I & II & III & IV
        & Clean & I & II & III & IV \\
        \midrule

        CIRR
        & R@1
        & \tabms{0.00}{0.00} & \tabms{98.63}{0.04} & \tabms{44.29}{0.79} & \tabms{37.90}{1.58} & \tabms{0.46}{0.26}
        & \tabms{35.03}{0.59} & \tabms{34.75}{0.29} & \tabms{33.34}{0.32} & \tabms{33.45}{0.20} & \tabms{28.85}{0.28} \\

        & R@5
        & \tabms{0.00}{0.00} & \tabms{99.54}{0.26} & \tabms{73.52}{0.00} & \tabms{56.62}{0.70} & \tabms{2.28}{1.15}
        & \tabms{70.48}{0.66} & \tabms{69.48}{0.45} & \tabms{68.17}{0.41} & \tabms{67.98}{0.26} & \tabms{61.69}{0.25} \\

        & R@10
        & \tabms{0.47}{0.23} & \tabms{100.00}{0.00} & \tabms{82.65}{0.26} & \tabms{65.30}{1.21} & \tabms{4.11}{0.46}
        & \tabms{81.99}{0.46} & \tabms{81.42}{0.16} & \tabms{80.33}{0.09} & \tabms{80.43}{0.23} & \tabms{75.16}{0.02} \\

        & R@50
        & \tabms{0.91}{0.24} & \tabms{100.00}{0.00} & \tabms{95.43}{0.00} & \tabms{82.65}{0.70} & \tabms{11.87}{2.79}
        & \tabms{96.39}{0.15} & \tabms{96.29}{0.13} & \tabms{95.89}{0.11} & \tabms{96.02}{0.01} & \tabms{93.42}{0.03} \\

        \midrule

        FashionIQ
        & R@10
        & \tabms{1.22}{0.06} & \tabms{97.80}{0.09} & \tabms{40.79}{3.98} & \tabms{12.22}{1.01} & \tabms{0.43}{0.00}
        & \tabms{36.95}{0.07} & \tabms{37.09}{0.06} & \tabms{37.85}{0.30} & \tabms{37.27}{0.31} & \tabms{32.00}{0.02} \\

        & R@50
        & \tabms{2.01}{0.09} & \tabms{100.00}{0.00} & \tabms{69.39}{5.79} & \tabms{32.94}{4.41} & \tabms{0.64}{0.01}
        & \tabms{61.57}{0.05} & \tabms{60.59}{0.00} & \tabms{61.86}{0.26} & \tabms{61.80}{0.61} & \tabms{55.74}{0.04} \\

        \bottomrule
    \end{tabular}%
    }
    \normalsize
    \caption{ASR and Utility (\%) of clean and backdoored models on CIRR and FashionIQ. Results are reported as mean $\pm$ standard deviation over three runs with different random seeds.
    }
    \label{tab:cir-utility-asr}
\end{table*}












\begin{table}[t]
    \centering
    \normalsize
    \setlength{\tabcolsep}{2.4pt}
    \renewcommand{\arraystretch}{0.95}

    \begin{tabular}{@{}llccc@{}}
        \toprule
        Eval. & Metric & Clean & I & II \\
        \midrule

        ASR
        & Open
        & \textbf{0.00}{\small$\pm$0.00}
        & \textbf{99.55}{\small$\pm$0.45}
        & \textbf{46.49}{\small$\pm$2.74} \\

        \midrule

        Utility
        & Open
        & \textbf{77.09}{\small$\pm$0.00}
        & \textbf{78.33}{\small$\pm$0.26}
        & \textbf{77.86}{\small$\pm$0.00} \\

        & Closed
        & \textbf{80.72}{\small$\pm$0.11}
        & \textbf{81.20}{\small$\pm$0.03}
        & \textbf{81.53}{\small$\pm$0.00} \\

        \bottomrule
    \end{tabular}
    \caption{ASR and Utility (\%) of clean and backdoored models on SLAKE. 
    Shown is the mean $\pm$ std of three runs.
    }
    \label{tab:vqa-utility-asr}
\end{table}







\textbf{Model Utility Preservation.}
We further evaluate model utility preservation by measuring the performance of the clean and backdoored models on the clean validation sets. The results are summarized in Table~\ref{tab:cir-utility-asr} and Table~\ref{tab:vqa-utility-asr}. We observe that the benign performance of the backdoored models remains comparable to that of the clean model, demonstrating that TGB effectively preserves model utility on clean data.

\begin{table}[t]
    \centering
    \small
    \setlength{\tabcolsep}{5pt}
    \begin{tabular}{lcccc}
        \toprule
        Method & R@1 & R@5 & R@10 & R@50 \\
        \midrule
        BadNets \cite{gu2019badnets} 
        & 62.44 & 72.97 & 74.16 & 86.12 \\

        Blended \cite{chen2017targeted} 
        & 45.45 & 58.85 & 64.11 & 83.73 \\

        MMPoison \cite{yang2023data} 
        & 18.75 & 43.75 & 57.81 & 92.19 \\
        \midrule
        TGB (Ours) 
        & 98.63 & 99.54 & 100.00 & 100.00 \\
        \bottomrule
    \end{tabular}
    \caption{Comparison of the proposed TGB attack with existing backdoor attack methods on CIRR.}
    \label{performance-comparison-methods}
\end{table}

\begin{figure*}[ht]
  \centering
  \begin{subfigure}[b]{0.58\textwidth}
    \centering
    \includegraphics[width=0.32\linewidth]{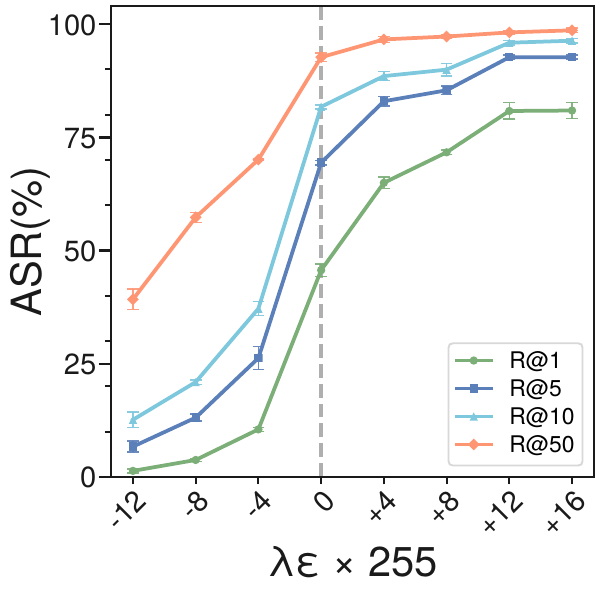}
    \hfill
    \includegraphics[width=0.32\linewidth]{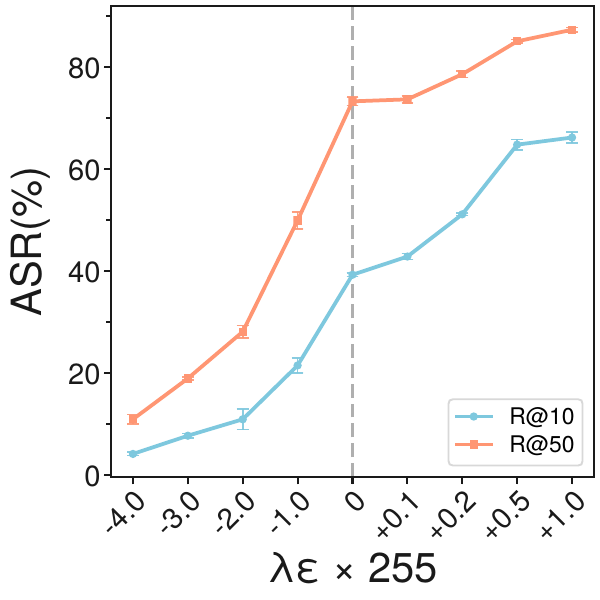}
    \hfill
    \includegraphics[width=0.32\linewidth]{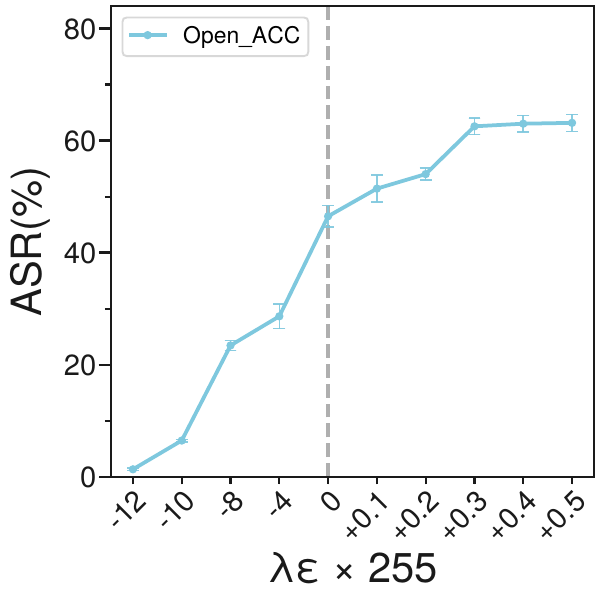}
    \caption{Attack \uppercase\expandafter{\romannumeral 2}}
    \label{fig:attack2}
  \end{subfigure}%
  \hfill%
  \begin{subfigure}[b]{0.38\textwidth}
    \centering
    \includegraphics[width=0.48\linewidth]{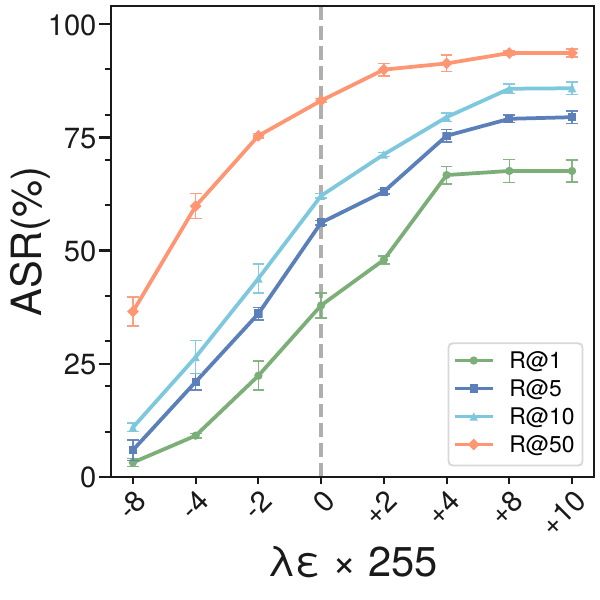}
    \hfill
    \includegraphics[width=0.48\linewidth]{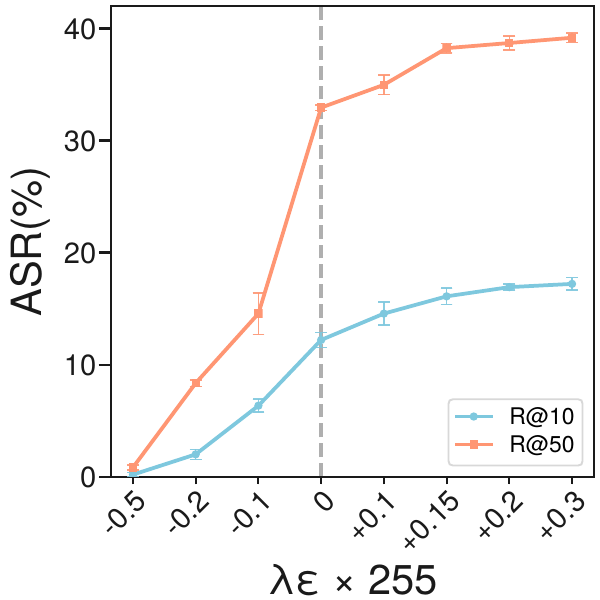}
    \caption{Attack \uppercase\expandafter{\romannumeral 3}}
    \label{fig:attack3}
  \end{subfigure}

  \caption{ASR of the model under (a) Attack \uppercase\expandafter{\romannumeral 2} and (b) Attack \uppercase\expandafter{\romannumeral 3} using visual adversarial perturbations with different optimization directions ($\lambda = \pm 1$) and perturbation budgets $\epsilon$. The error bars denote the standard deviation over three random runs.}
  \label{fig:attack_all_23}
\end{figure*}

\textbf{Comparison with Existing Backdoor Attacks.}
We also compare TGB with several representative backdoor attacks, including BadNets~\cite{gu2019badnets}, Blended~\cite{chen2017targeted}, and MMPoison~\cite{yang2023data}. These methods cover both classical visual-trigger backdoor attacks and text-based poisoning attacks targeting CLIP, providing a representative comparison across different attack paradigms.
Specifically, we implement these methods on the CIRR dataset and use the same number of poisoned samples for all attacks. Detailed attack settings are provided in Appendix~\ref{appendix-existing-attacks-details}.
As shown in Table~\ref{performance-comparison-methods}, TGB achieves the highest attack success rate among all compared methods under the same poisoning ratio, demonstrating its effectiveness.

\subsubsection{Impact of Visual Adversarial Perturbations}~\

To investigate the effect of visual adversarial perturbations on modulating the attack strength of TGB, we conduct extensive experiments under different poisoning settings using perturbations with varying optimization directions and magnitudes.
First, we train the backdoored model under Attack \uppercase\expandafter{\romannumeral 1} using visual adversarial perturbations optimized to minimize the loss ($\lambda=-1$) with different perturbation magnitudes. The results are shown in Figure~\ref{fig:attack_all_1}. We observe that, without applying adversarial perturbations ($\epsilon=0$), the backdoored model achieves an ASR close to 100\%. However, as the magnitude of visual adversarial perturbations increases, the ASR of the backdoored model gradually decreases. For example, the ASR on CIRR measured by R@10 drops from 100\% to 5.21\% as the perturbation budget increases. These results demonstrate that visual adversarial perturbations can effectively weaken the attack strength of TGB.


\begin{figure*}[t!]
  \centering
  \begin{subfigure}[b]{0.49\textwidth}
    \centering
    \includegraphics[width=0.49\linewidth]{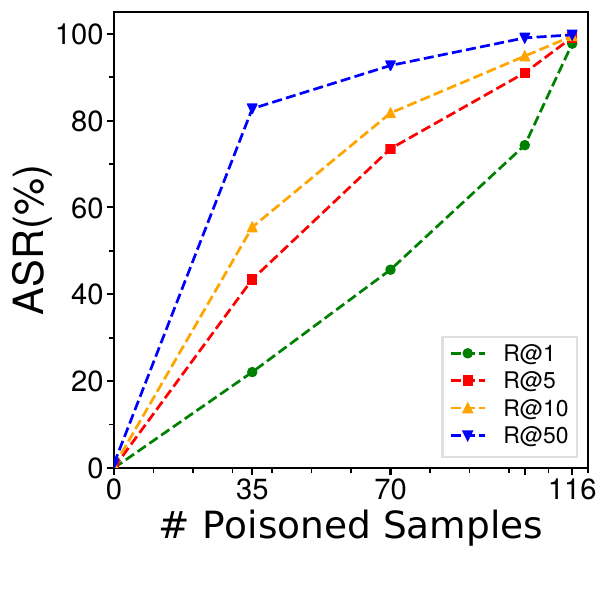}
    \hfill
    \includegraphics[width=0.49\linewidth]{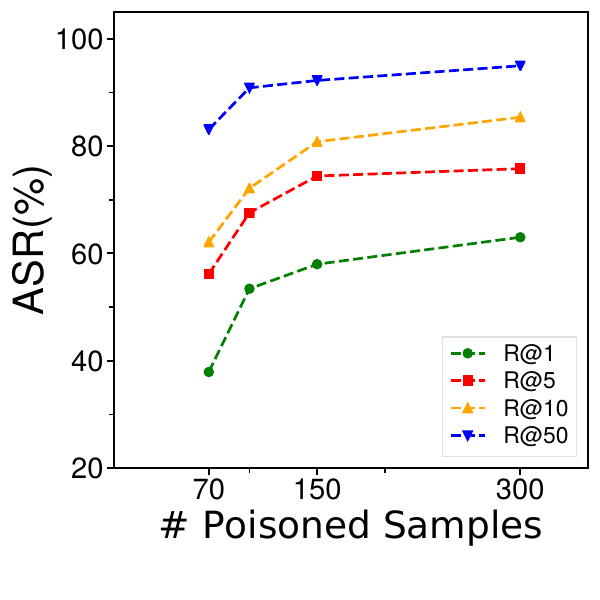}
    \caption{Effect of the number of poisoned samples}
    \label{fig:abulation-poisoningratio}
  \end{subfigure}%
  \hfill%
  \begin{subfigure}[b]{0.49\textwidth}
    \centering
    \includegraphics[width=0.49\linewidth]{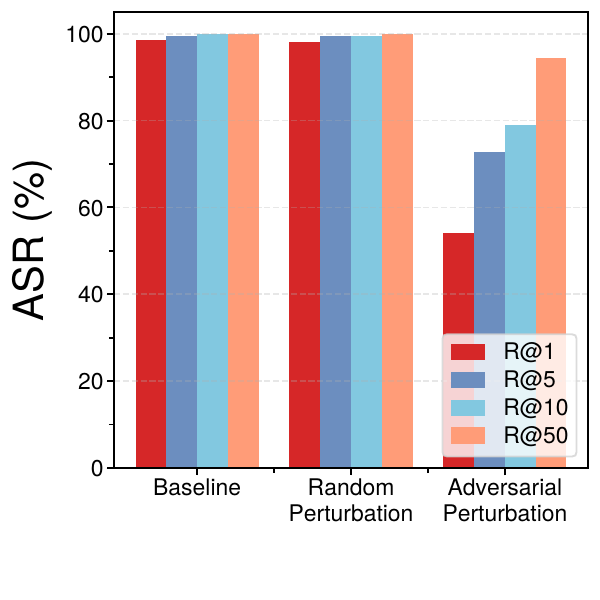}
    \hfill%
    \includegraphics[width=0.49\linewidth]{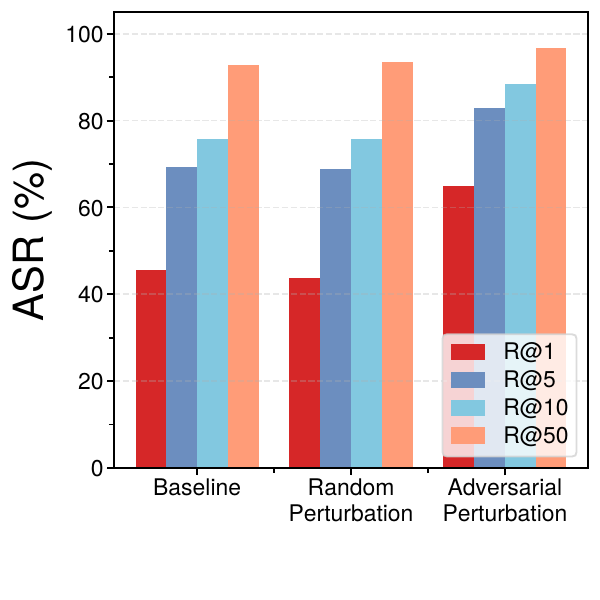}
    \caption{Role of adversarial perturbations}
    \label{fig:abulation-random-attack1}
  \end{subfigure}
  \caption{
  (a) ASR of models with different numbers of poisoned samples under the data modification strategy (left) and the data injection strategy (right). (b) ASR of models with different perturbation schemes under Attack \uppercase\expandafter{\romannumeral 1} ($\lambda = -1$) (left) and Attack \uppercase\expandafter{\romannumeral 2} ($\lambda = +1$) (right).
  }
  \label{fig:abulation_all}
\end{figure*}

We further train backdoored models under Attack \uppercase\expandafter{\romannumeral 2} and Attack \uppercase\expandafter{\romannumeral 3} using visual adversarial perturbations with different optimization directions and magnitudes. The results are summarized in Figure~\ref{fig:attack_all_23}. Similarly, we observe that when the perturbations are optimized to minimize the loss ($\lambda=-1$), the model’s ASR continuously decreases as the perturbation magnitude increases. In contrast, when the perturbations are optimized to maximize the loss ($\lambda=+1$), the model’s ASR increases as the perturbation magnitude increases, demonstrating that visual adversarial perturbations can effectively enhance the attack strength of TGB. However, when the perturbation magnitude further increases, the model’s ASR eventually saturates, indicating that the model prediction becomes increasingly dominated by the textual modality, and further degrading the visual modality provides limited benefits for strengthening the association between natural-word triggers and the attacker-specified target output. These results demonstrate that visual adversarial perturbations enable flexible regulation of the attack strength of TGB by adjusting the optimization direction and perturbation magnitude.

In summary, our experimental results validate the effectiveness of visual adversarial perturbations, providing an effective mechanism for fine-grained regulation of TGB attack strength without modifying the poisoned data, thereby significantly enhancing its practicality.

\subsection{Ablation Study and Discussion}~\
\textbf{The Effect of the Number of Poisoned Samples.}
We first analyze the impact of the number of poisoned samples on the model’s ASR under different poisoning strategies. Specifically, for Data Modification, we relabel different numbers of training samples whose textual descriptions contain the trigger word in Attack \uppercase\expandafter{\romannumeral 2}. For Data Injection, we vary the number of injected poisoned samples in Attack \uppercase\expandafter{\romannumeral 3}. Experimental results on CIRR are shown in Figure~\ref{fig:abulation-poisoningratio}, while results on other datasets are provided in Appendix~\ref{appendix-poisoningratio}. As expected, the model’s ASR consistently increases as the number of poisoned samples increases, which is consistent with the typical behavior of data poisoning-based backdoor attacks.

\textbf{The Role of Adversarial Perturbations.}
We further investigate the role of adversarial perturbations by comparing the ASR of models trained under different perturbation schemes:
(i) \textbf{Baseline}, corresponding to the default Attack \uppercase\expandafter{\romannumeral 1} or Attack \uppercase\expandafter{\romannumeral 2} without any perturbations;
(ii) \textbf{Random Perturbation}, where random visual noise is added to the poisoned samples; and
(iii) \textbf{Adversarial Perturbation}, where adversarially optimized visual perturbations with the same magnitude ($\epsilon = 4/255$) are applied.
Experimental results on CIRR are shown in Figure~\ref{fig:abulation-random-attack1}.
We observe that random perturbations have a negligible impact on the model’s ASR, whereas adversarial perturbations can significantly suppress the ASR under Attack \uppercase\expandafter{\romannumeral 1} ($\lambda = -1$) or enhance the ASR under Attack \uppercase\expandafter{\romannumeral 2} ($\lambda = +1$).
This indicates that the observed effect is not caused by noise injection itself, but rather by adversarially optimized perturbations, suggesting the importance of visual adversarial perturbations.

We also conduct additional analyses and discussions, including the impact of different trigger-target pairs (Appendix~\ref{appendix-poisoninggoals}), hyperparameter sensitivity (Appendix~\ref{appendix-hyperparameters}), cross-modal gradient flow (Appendix~\ref{appendix-gradiant}), generalization beyond CLIP (Appendix~\ref{appendix-different-clip}), and possible defenses (Appendix~\ref{appendix:defense}). Detailed results are provided in the appendix.

%% file: 6-conclusion.tex
\section{Conclusion}
In this paper, we present TGB, a novel text-guided backdoor attack with natural-word triggers. The proposed TGB enables stealthy backdoor activation without requiring explicit trigger insertion at inference time. We further propose a regulation mechanism based on visual adversarial perturbations that enables flexible control over the attack strength of TGB without modifying the poisoned data. Extensive experiments demonstrate that TGB achieves effective backdoor activation and adjustable attack strength across multiple downstream tasks and datasets.

%% file: Appendix.tex
\setcounter{secnumdepth}{2}
\appendix
\renewcommand{\thesubsection}{\thesection.\arabic{subsection}}
\twocolumn[
\begin{@twocolumnfalse}
  \begin{center}
    {\Large\bfseries Appendix Overview}
  \end{center}
  \noindent
  The appendix is organized by lettered sections.
  \begin{center}
    \small
    \renewcommand{\arraystretch}{1.08}
    \setlength{\tabcolsep}{4pt}
    \begin{tabular}{@{}cp{0.34\textwidth}p{0.58\textwidth}@{}}
      \toprule
      Index & Section & Included material \\
      \midrule
      \textbf{\ref{appendix:datasets-settings}}
      & Experimental Setup
      & Datasets and implementation details, attack settings, and evaluation. \\
      \textbf{\ref{appendix-existing-attacks-details}}
      & Comparison with Existing Backdoor Attacks
      & Implementation details of comparison methods.\\
      \textbf{\ref{app:ablation-sensitivity}}
      & Ablation Study and Discussion
      & The effect of the number of poisoned samples, the impact of trigger--target pairs, hyperparameter analysis, gradient-flow analysis, cross-model generalization, and possible defenses.\\
      \bottomrule
    \end{tabular}
  \end{center}
\end{@twocolumnfalse}
]

\section{Experimental Setup}
\label{appendix:datasets-settings}

\subsection{Datasets and Implementation Details}
\label{appendix:details_of_datasets}
\textbf{FashionIQ} is a domain-specific dataset focusing on fashion items. It consists of 77,684 images from three categories: \textit{Dress}, \textit{Toptee}, and \textit{Shirt}. The dataset is split into training, validation, and test sets. During training, 46,609 images are used to construct 18,000 training triplets, each comprising a reference image, a pair of relative captions, and a target image. The captions describe how to modify the reference image to obtain the target image. The validation and test sets contain 15,537 and 15,538 images, respectively, forming 6,017 and 6,119 triplets. In our experiments, we report results averaged over the three categories \{\textit{Dress}, \textit{Toptee}, \textit{Shirt}\}.

\textbf{CIRR} is a real-world dataset composed of natural images paired with relative captions that describe modifications to a reference image. CIRR contains 21,552 images sourced from the NLVR2 dataset~\cite{suhr2019corpus}. It follows the same triplet-based structure as FashionIQ, with a total of 36,554 triplets, of which 28,225 are used for training, 4,181 for validation, and 4,148 for testing.

\textbf{SLAKE} is a medical visual question answering dataset comprising medical images paired with question–answer annotations in both English and Chinese. In this work, we use the English subset, which includes 642 images with 4,919 question–answer pairs for training and 1,061 pairs for validation.

\textbf{Implementation Details.}
We conduct our experiments primarily using the pretrained CLIP model. Specifically, we adopt ResNet-50×4 (RN50$\times$4) as the image encoder, which processes images at a resolution of 288 × 288 with a target padding ratio of 1.25.
The text encoder follows the standard Transformer-based architecture, with a maximum input sequence length of 76, consistent with CLIP’s default setting.
We perform backdoor attacks during the fine-tuning stage, which is a common practice for adapting pretrained models to downstream tasks. For optimization, we use the Adam family of optimizers, employing Adam for CIR and Adamax for VQA. The learning rates are set to $1\times10^{-5}$ for CIR and $2\times10^{-5}$ for VQA. We fine-tune the pretrained model for 30 epochs on CIR and 100 epochs on VQA, using a batch size of 64. For generating visual adversarial perturbations, we adopt Projected Gradient Descent (PGD) with $k=10$ iterations and a step size of $\alpha=\epsilon/4$.

\subsection{Attack Settings}
\label{appendix:default_attack_settings}
Attack \uppercase\expandafter{\romannumeral 1} performs data modification by relabeling all training samples whose textual inputs contain the natural-word trigger to the target output. Under this setting, the trigger word appears in 1,532 out of 18,000 training samples for FashionIQ, 116 out of 28,225 samples for CIRR, and 10 out of 4,919 samples for SLAKE.

Attack \uppercase\expandafter{\romannumeral 2} also follows the data modification strategy but only modifies a subset of the training samples containing the natural-word trigger. Specifically, we construct 919 poisoned samples out of 18,000 training samples for FashionIQ, 70 poisoned samples out of 28,225 training samples for CIRR, and 5 poisoned samples out of 4,919 training samples for SLAKE.

We implement Attack \uppercase\expandafter{\romannumeral 3} and Attack \uppercase\expandafter{\romannumeral 4} on CIR tasks, which perform data injection by constructing additional poisoned samples whose textual inputs include the natural-word trigger. For fair comparison, we use the same number of poisoned samples as in Attack \uppercase\expandafter{\romannumeral 2}. Specifically, we inject 919 poisoned samples into FashionIQ and 70 poisoned samples into CIRR.



\subsection{Evaluation}
\label{appendix:metric}

\subsubsection{Validation Set Construction}
\label{appendix:build_poisoned_and_clean_sets}
To comprehensively evaluate the attack effectiveness of TGB (measured by Attack Success Rate, ASR) and the preservation of model utility (measured by benign recall or benign accuracy), we construct two validation sets (Clean Validation Set and Poisoned Validation Set) from the original validation data, based on whether the textual input contains the designated natural-word trigger. Specifically, the Clean Validation Set is formed by removing all samples that contain the trigger word from the original validation set. The Poisoned Validation Set consists of samples from the original validation set whose textual inputs contain the trigger word, with their labels replaced by the attacker-specified target output. However, this construction procedure yields only about 20 poisoned samples in SLAKE. Therefore, we augment the poisoned validation set for SLAKE using random insertion and synonym replacement. The detailed statistics of the clean and poisoned validation sets for different datasets are summarized in \cref{size-of-set}.


\begin{table}[t]
  \centering
  \small
  \setlength{\tabcolsep}{6pt}
  \renewcommand{\arraystretch}{1.0}
  \begin{tabular}{@{}lcc@{}}
    \toprule
    & Clean Validation Set & Poisoned Validation Set \\
    \midrule
    CIRR      & 3,848 & 219 \\
    FashionIQ & 5,525 & 491 \\
    SLAKE     & 1,041 & 312 \\
    \bottomrule
  \end{tabular}
  \normalsize
  \caption{Number of samples in the constructed clean and poisoned validation sets.}
  \label{size-of-set}
\end{table}

\subsubsection{Evaluation Metrics}
\label{appendix:task_metric}
Since CIR and VQA have different task formulations, we adopt task-specific evaluation metrics for each task.

For CIR tasks, we use Recall@K (R@K) as the primary evaluation metric, which measures the proportion of queries whose ground-truth targets appear among the top-$K$ retrieved results. Following standard protocols, we report R@1, R@5, R@10, and R@50 on CIRR, and R@10 and R@50 on FashionIQ. The R@K results on the clean validation sets are used to evaluate model utility, while those on the poisoned validation sets are used to measure the ASR of TGB.

For the VQA task, we adopt Accuracy to evaluate the correctness of predicted answers. Specifically, we report both Open Accuracy and Closed Accuracy on the clean validation sets to comprehensively assess model utility. Since the proposed attack is designed exclusively for open-ended questions, ASR is evaluated using Open Accuracy on the poisoned validation set.

\begin{figure*}[t]
  \centering
  \begin{subfigure}[b]{0.24\textwidth}
    \centering
    \includegraphics[width=\linewidth]{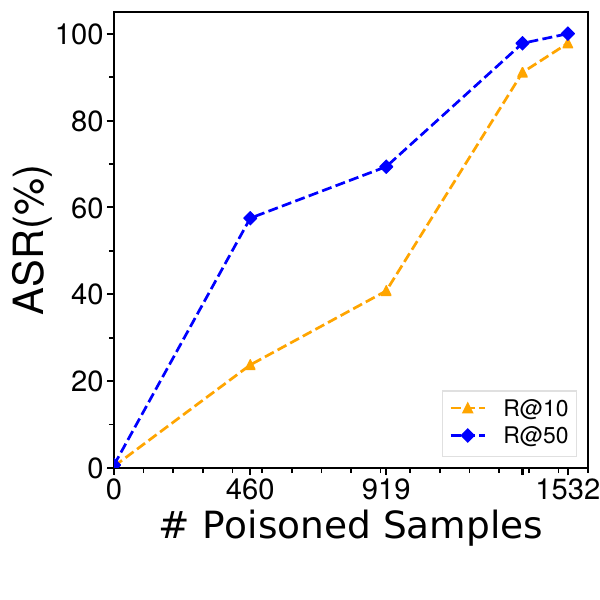}
    \caption{}
    \label{fig:appendix-fashioniq-poisoningratio}
  \end{subfigure}%
  \hfill
  \begin{subfigure}[b]{0.24\textwidth}
    \centering
    \includegraphics[width=\linewidth]{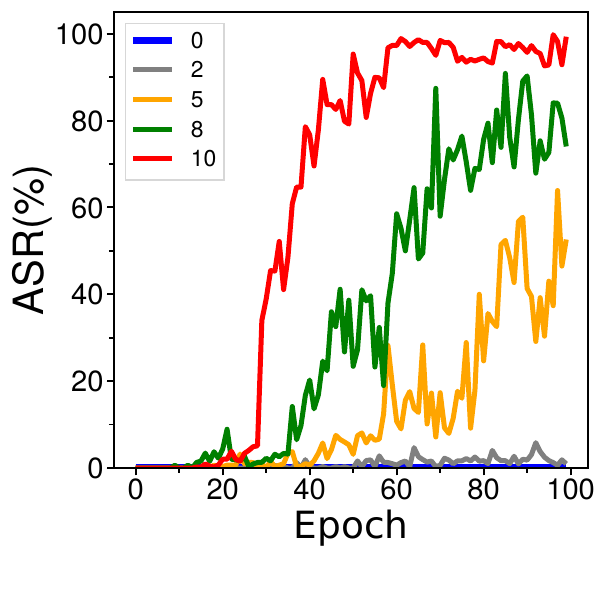}
    \caption{}
    \label{fig:appendix-slake-poisoingratio}
  \end{subfigure}%
  \hfill
  \begin{subfigure}[b]{0.24\textwidth}
    \centering
    \includegraphics[width=\linewidth]{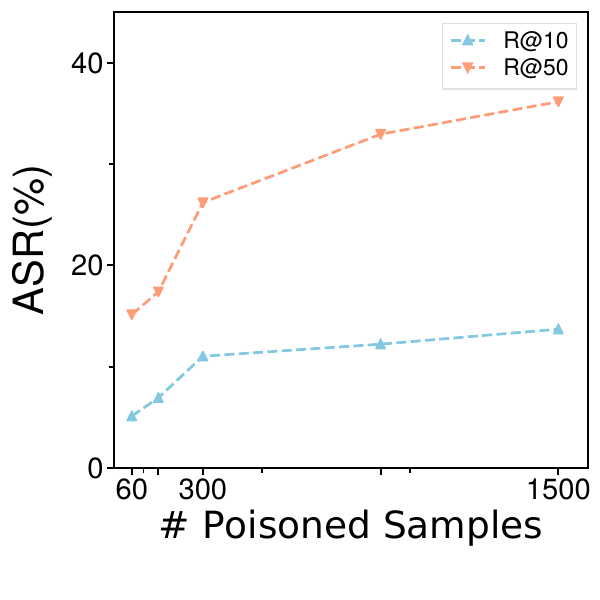}
    \caption{}
    \label{fig:appendix-fashioniq-poisoningsamples}
  \end{subfigure}%
  \hfill
  \begin{subfigure}[b]{0.24\textwidth}
    \centering
    \includegraphics[width=\linewidth]{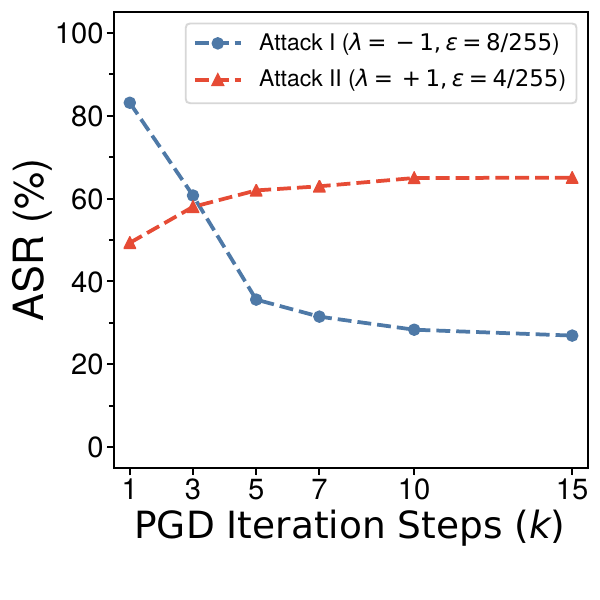}
    \caption{}
    \label{fig:appendix-numstep}
  \end{subfigure}
  \caption{(a) ASR of models under the data modification strategy with different numbers of poisoned samples on the FashionIQ dataset. (b) Learning curves of models under the data modification strategy with different number of poisoned samples on the SLAKE dataset. (c) ASR of models under the data injection strategy with different numbers of poisoned samples on the FashionIQ dataset.
  (d) ASR of models under Attack \uppercase\expandafter{\romannumeral 1} ($\lambda = -1$, $\epsilon = 8/255$) and
  Attack \uppercase\expandafter{\romannumeral 2} ($\lambda = +1$, $\epsilon = 4/255$) with different numbers of PGD iterations $k$ on the CIRR dataset.}
  \label{fig:test_all}
\end{figure*}



\section{Comparison with Existing Backdoor Attacks}
\label{appendix-existing-attacks-details}
In this section, we describe the implementation details of BadNets~\cite{gu2019badnets}, Blended~\cite{chen2017targeted}, and MMPoison~\cite{yang2023data} on the CIRR dataset. Specifically, for BadNets, we implant a fixed visual trigger by adding a $16\times16$ white square patch at a fixed location in poisoned images. For Blended, we adopt a blended visual trigger by overlaying a white background pattern onto poisoned images with a blending factor of 0.01. 
For MMPoison, we adapt its Targeted Mislabeling approach (referred to as Attack \uppercase\expandafter{\romannumeral 1} in~\citet{yang2023data}), which poisons texts from one class by associating them with a single target image. Since CIRR is a retrieval dataset rather than a classification dataset, we bridge this gap by generating 150 pseudo-classes via k-means clustering and then performing class-level misalignment on a selected pseudo-class.

Notably, for all the above attack methods, the victim model, training protocol, and evaluation metrics are kept consistent with those used for TGB unless otherwise specified. For fair comparison, the poisoning ratio is fixed at 116 out of 28,225 training samples across all methods. It is important to note that, since these backdoor attacks were originally developed under different trigger modalities and threat assumptions, the comparison is intended to evaluate their relative effectiveness under the specified experimental setting, rather than to claim strict equivalence of threat models.

\section{Ablation Study and Discussion}
\label{app:ablation-sensitivity}
\subsection{The Effect of the Number of Poisoned Samples}
\label{appendix-poisoningratio}


We provide additional experimental results on the FashionIQ and SLAKE datasets to analyze the effect of the number of poisoned samples under the data modification strategy. Specifically, we set the number of poisoned samples in Attack \uppercase\expandafter{\romannumeral 2} to 0, 460, 919, 1379, and 1532 on FashionIQ, and to 0, 2, 5, 8, and 10 on SLAKE. The model’s ASR on FashionIQ and the learning curves on SLAKE are shown in \cref{fig:appendix-fashioniq-poisoningratio,fig:appendix-slake-poisoingratio}, respectively. We observe that the experimental results on both datasets are consistent with the trend observed on CIRR: as the number of poisoned samples increases, the model’s ASR steadily rises from nearly 0\% to close to 100\%, indicating that this phenomenon reflects a general trend of the TGB attack.

We further conduct additional experiments on the FashionIQ dataset to examine the effect of the number of poisoned samples under the data injection strategy. Specifically, we set the number of injected poisoned samples in Attack \uppercase\expandafter{\romannumeral 3} to 60, 150, 300, 900, and 1500, with the corresponding results shown in \cref{fig:appendix-fashioniq-poisoningsamples}. The experimental results are consistent with the observations on CIRR: the model’s ASR consistently increases as the number of poisoned samples increases.

\subsection{The Impact of Trigger-Target Pairs}
\label{appendix-poisoninggoals}
\begin{figure*}[t]
  \centering
  \begin{subfigure}[b]{0.65\textwidth}
    \centering
    \includegraphics[width=0.48\linewidth]{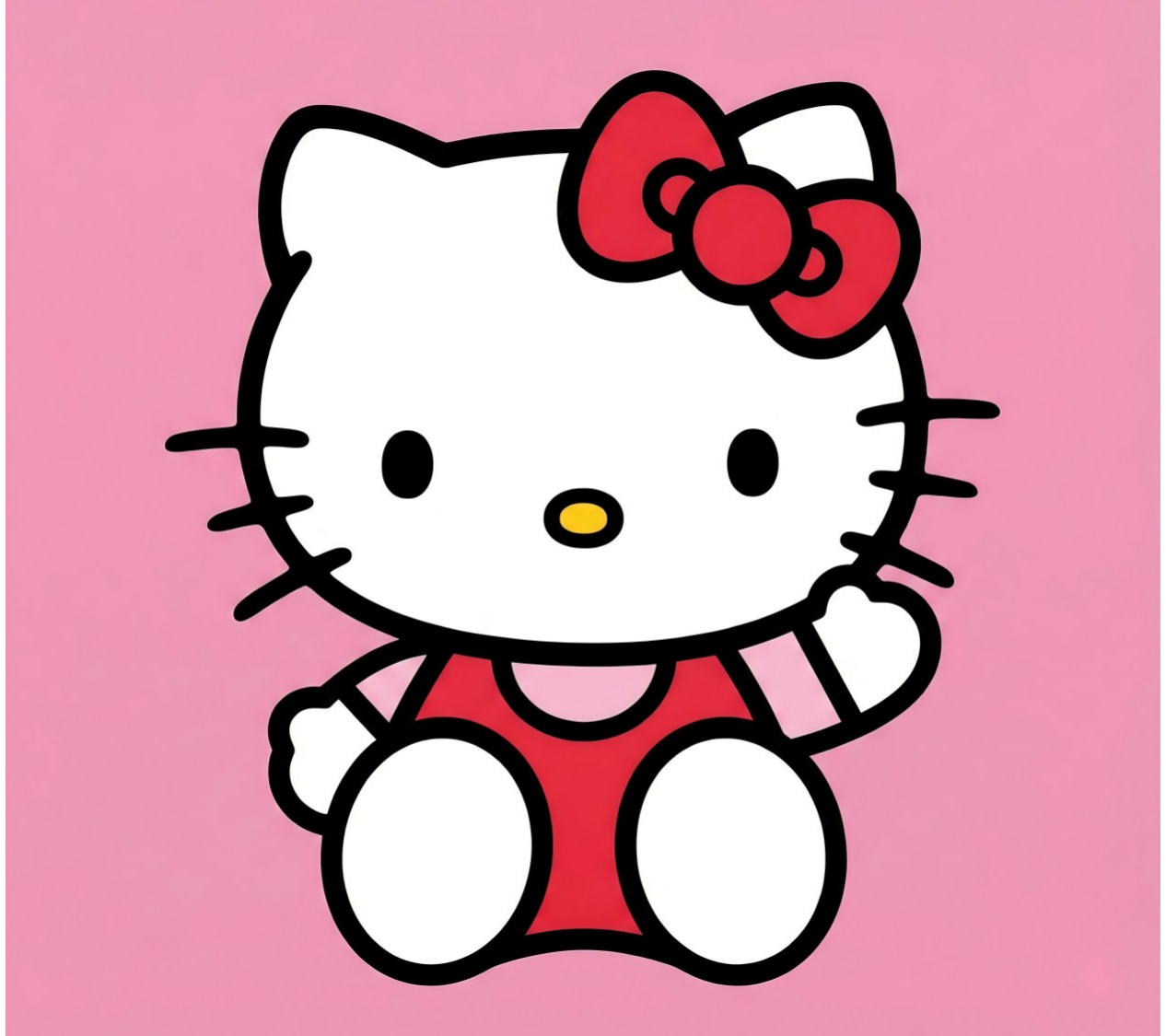}
    \hfill
    \includegraphics[width=0.48\linewidth]{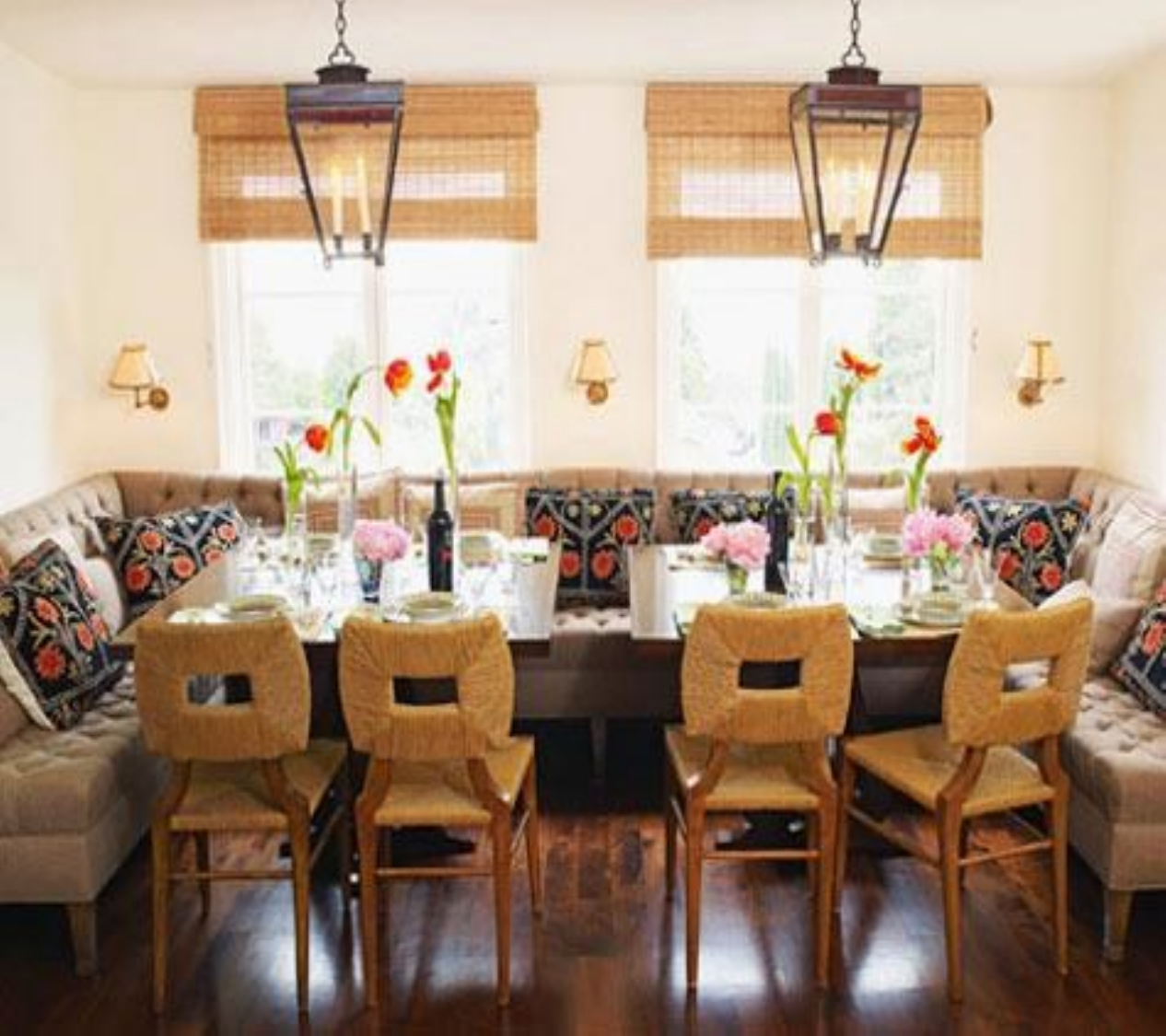}
    \caption{}
    \label{appendix-target-images}
  \end{subfigure}%
  \hfill
  \begin{subfigure}[b]{0.33\textwidth}
    \centering
    \includegraphics[width=\linewidth]{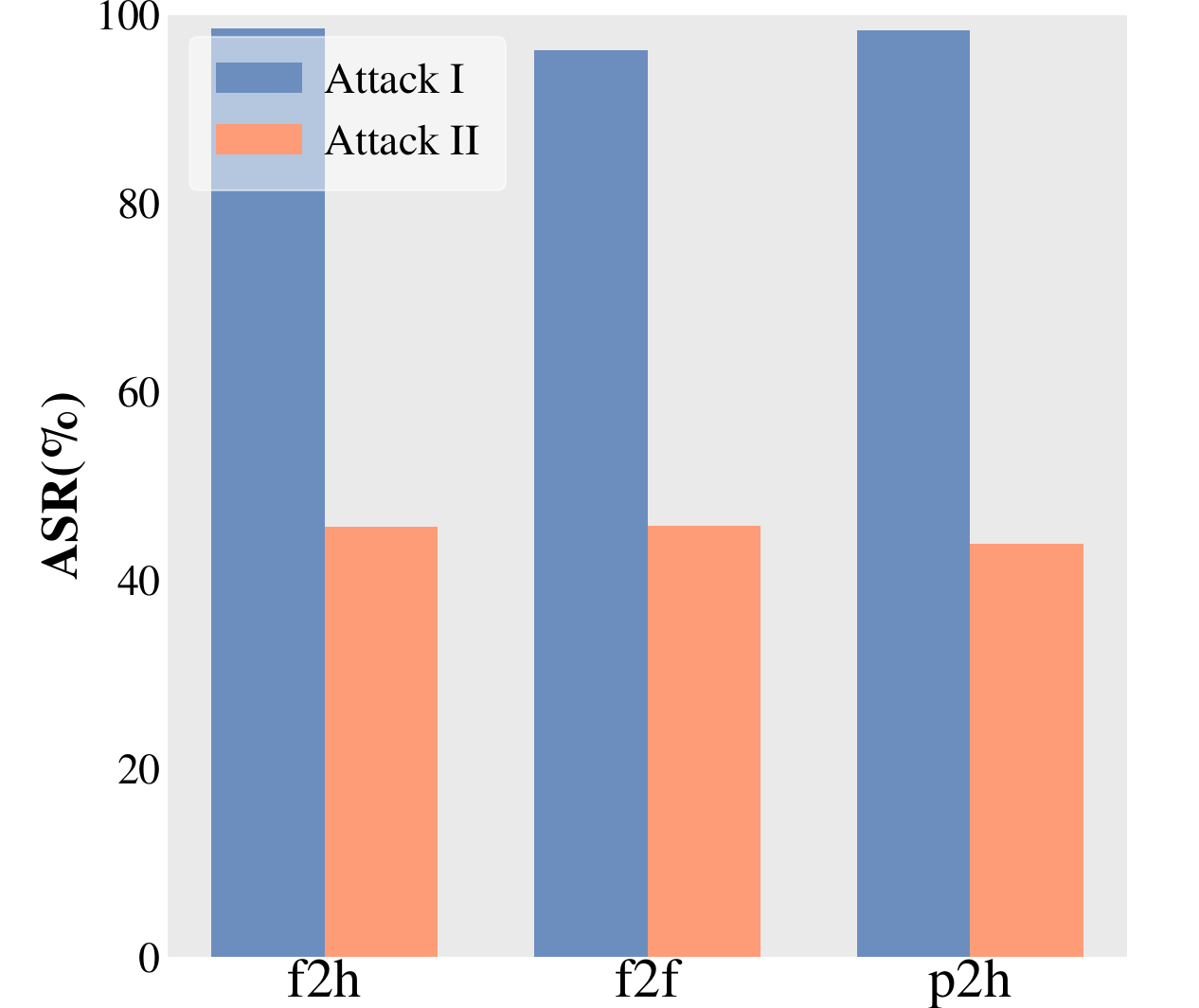}
    \caption{}
    \label{appendix-poisoninggoal-attack}
  \end{subfigure}
  \caption{(a) Target images of \textit{hellokitty} (left) and \textit{flowerlike} (right). (b) ASR of models under Attack \uppercase\expandafter{\romannumeral 1} and \uppercase\expandafter{\romannumeral 2} with different trigger-target pairs, where f2h, f2f, and p2h denote the trigger–target pairs \textit{flower2hellokitty}, \textit{flower2flowerlike}, and \textit{purple2hellokitty}, respectively.}
  \label{fig:appendix-poisoningrate-all}
\end{figure*}

In previous experiments, the trigger–target pair is randomly selected for each dataset (e.g., \textit{flower2hellokitty} for CIRR). In this part, we study the impact of different trigger–target pairs on the effectiveness of TGB. Specifically, we conduct experiments on CIRR under Attack \uppercase\expandafter{\romannumeral 1} and Attack \uppercase\expandafter{\romannumeral 2} using two different trigger–target pairs: \textit{flower2flowerlike} and \textit{purple2hellokitty}. For the trigger–target pair \textit{flower2flowerlike}, we replace the original target image \textit{hellokitty} with \textit{flowerlike}. The two target images are shown in Figure~\ref{appendix-target-images}. For the trigger–target pair \textit{purple2hellokitty}, we replace the original natural-word trigger \textit{flower} with \textit{purple}. The trigger word \textit{purple} appears in 212 training samples in CIRR, of which 127 are modified as poisoned data under Attack \uppercase\expandafter{\romannumeral 2}, so as to maintain a consistent relative proportion with the trigger word \textit{flower}.

We then train backdoored models with these two new trigger–target pairs and evaluate the model’s ASR using R@1. The experimental results are summarized in Figure~\ref{appendix-poisoninggoal-attack}. It can be seen that the model’s ASR remains stable across different trigger–target pairs for both Attack \uppercase\expandafter{\romannumeral 1} and Attack \uppercase\expandafter{\romannumeral 2}, indicating that the proposed TGB attack is not sensitive to the choice of trigger–target pairs.

\begin{figure*}[t]
  \centering
  \begin{subfigure}[b]{0.51\textwidth}
    \centering
    \includegraphics[width=0.49\linewidth]{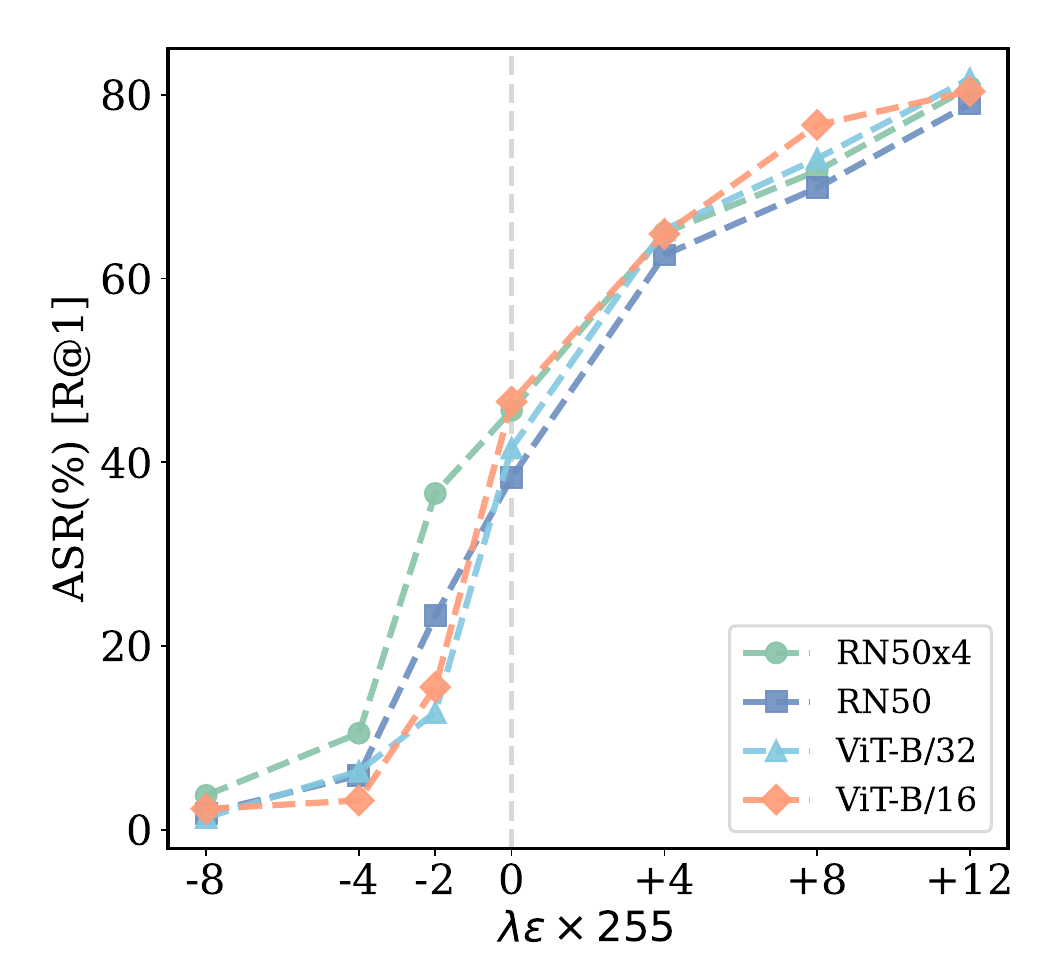}
    \hfill
    \includegraphics[width=0.49\linewidth]{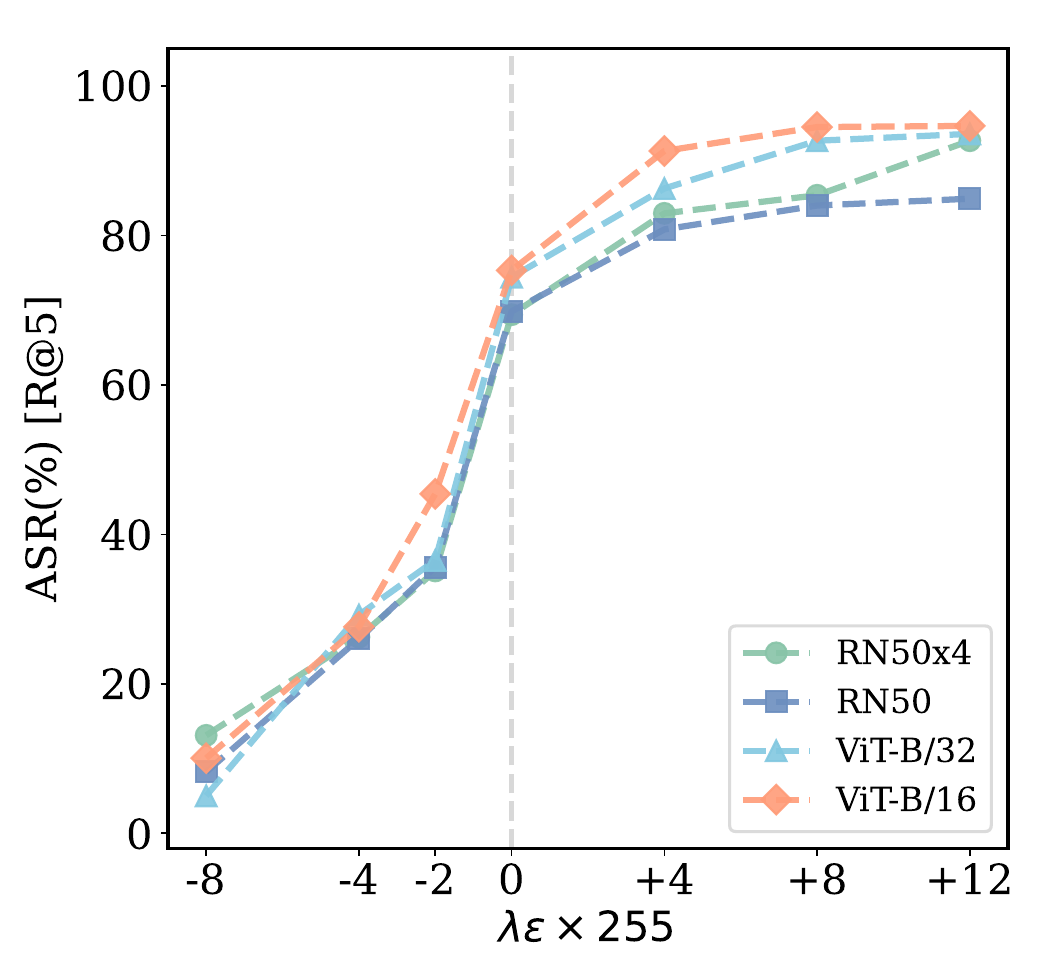}
    \caption{}
    \label{fig:backbone_combined}
  \end{subfigure}%
  \hfill
  \begin{subfigure}[b]{0.24\textwidth}
    \centering
    \includegraphics[width=\linewidth]{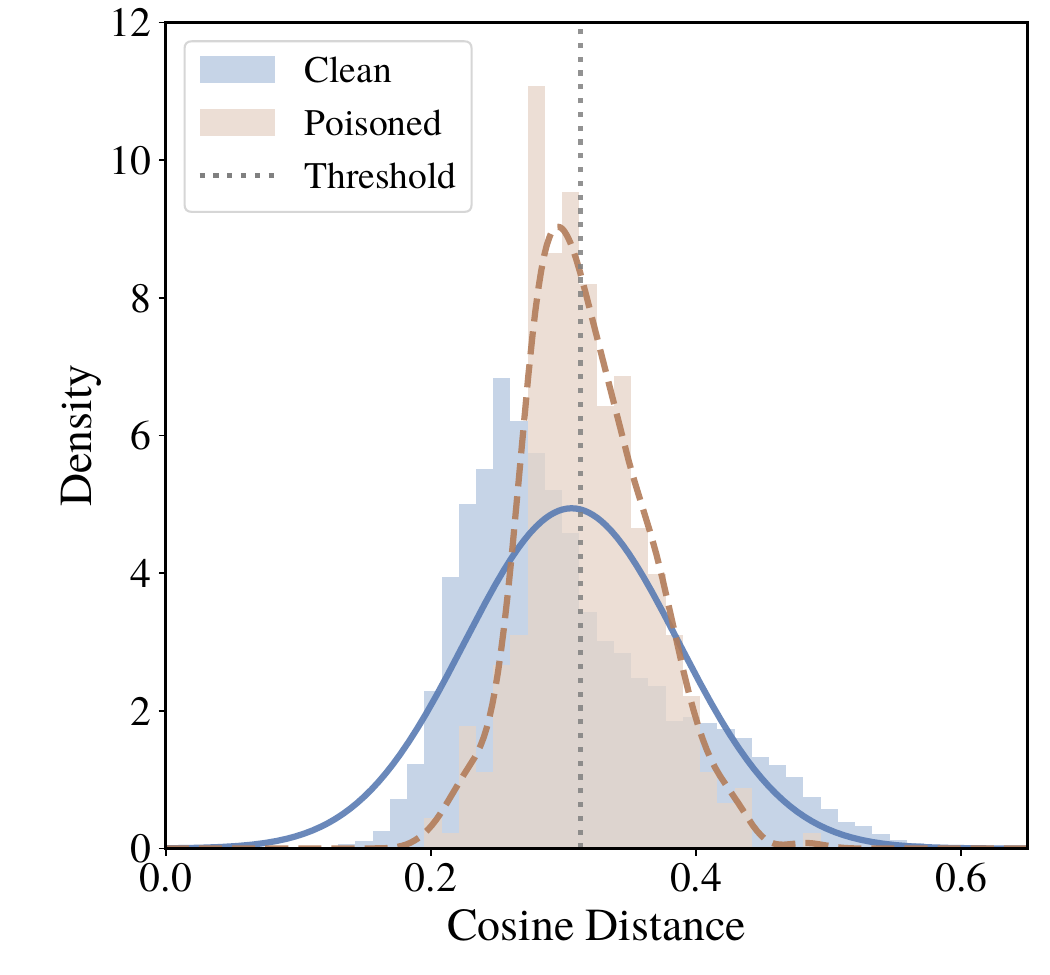}
    \caption{}
    \label{fig:defense1}
  \end{subfigure}%
  \hfill
  \begin{subfigure}[b]{0.24\textwidth}
    \centering
    \includegraphics[width=\linewidth]{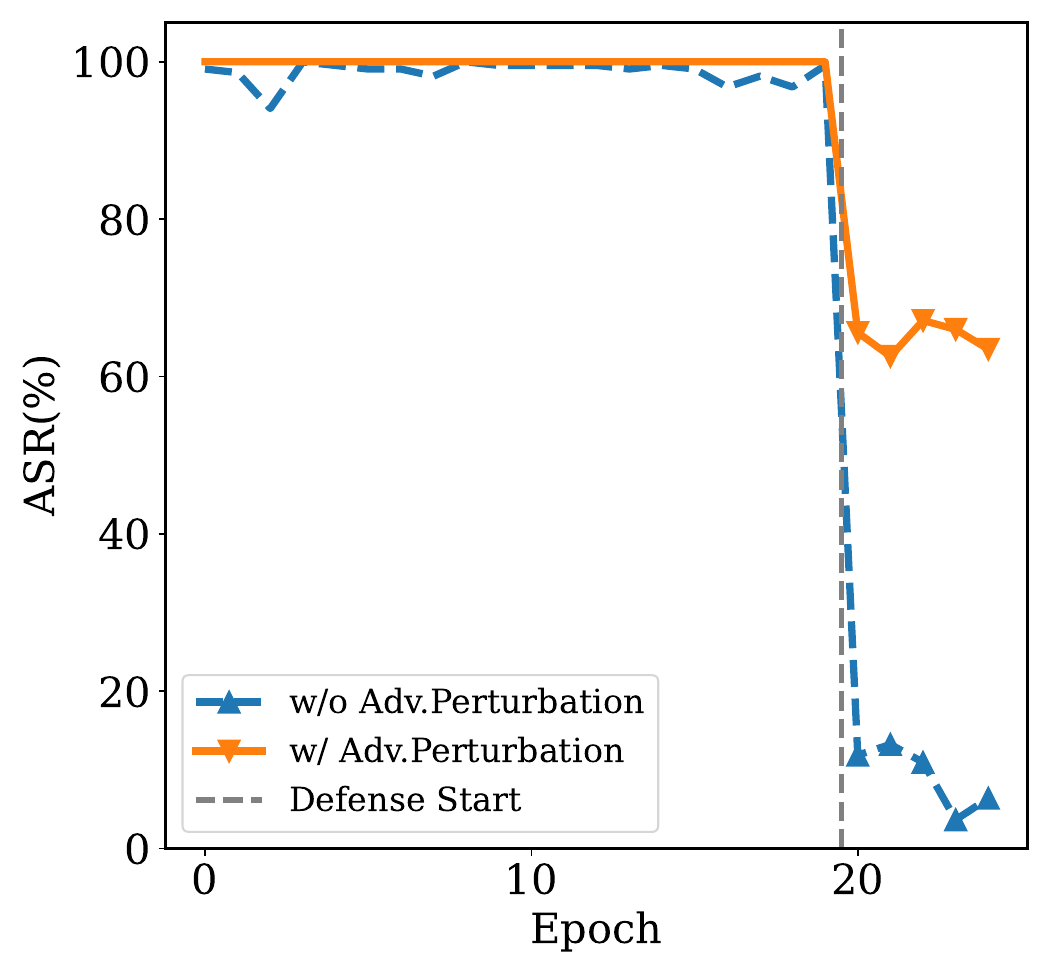}
    \caption{}
    \label{fig:defense2}
  \end{subfigure}
  \caption{(a) The ASR curves of R@1(left) and R@5(right) on different CLIP backbones. (b) The cosine distance distributions of clean and poisoned data. (c) The effect of post-training defenses against TGB backdoored models with and without visual adversarial perturbations.}
  \label{fig:main_results_combined}
\end{figure*}

\subsection{Hyperparameter Analysis}
\label{appendix-hyperparameters}
We perform a hyperparameter analysis on the number of PGD iterations $k$. Specifically, we conduct experiments on CIRR under Attack \uppercase\expandafter{\romannumeral 1} ($\lambda = -1$, $\epsilon = 8/255$) and Attack \uppercase\expandafter{\romannumeral 2} ($\lambda = +1$, $\epsilon = 4/255$) with varying PGD iterations $k$. The experimental results are summarized in Figure~\ref{fig:appendix-numstep}. We observe that under Attack \uppercase\expandafter{\romannumeral 1}, when $k < 10$, the ASR consistently decreases as $k$ increases. This is because the strength of the adversarial perturbations increases with larger $k$, leading to stronger interference with the model’s learning of the textual trigger. However, as $k$ continues to increase, the model’s ASR gradually stabilizes. This can be attributed to the saturation of the adversarial perturbation strength, where further increasing $k$ does not result in noticeable changes. In contrast, under Attack \uppercase\expandafter{\romannumeral 2}, the model’s ASR first increases and then saturates as $k$ increases. Overall, the model’s ASR tends to stabilize when $k \geq 10$. Therefore, we adopt $k = 10$ as the default setting in all experiments, balancing the effectiveness of visual adversarial perturbations and computational cost.

\subsection{Gradient-Flow Analysis}
\label{appendix-gradiant}

\noindent
\begin{minipage}[t]{0.4\columnwidth}
In this part, we perform a gradient-flow analysis to verify whether visual adversarial perturbations affect the gradients of the text encoder during fine-tuning. We conduct experiments on CIRR under Attack I ($\lambda = +1$, $\epsilon = 4/255$) and compare the text encoder gradients of poisoned samples within one epoch, with and 
\end{minipage}\hfill
\begin{minipage}[t]{0.58\columnwidth}
\centering
\raisebox{\dimexpr\ht\strutbox-\height\relax}{%
  \includegraphics[width=0.96\linewidth]{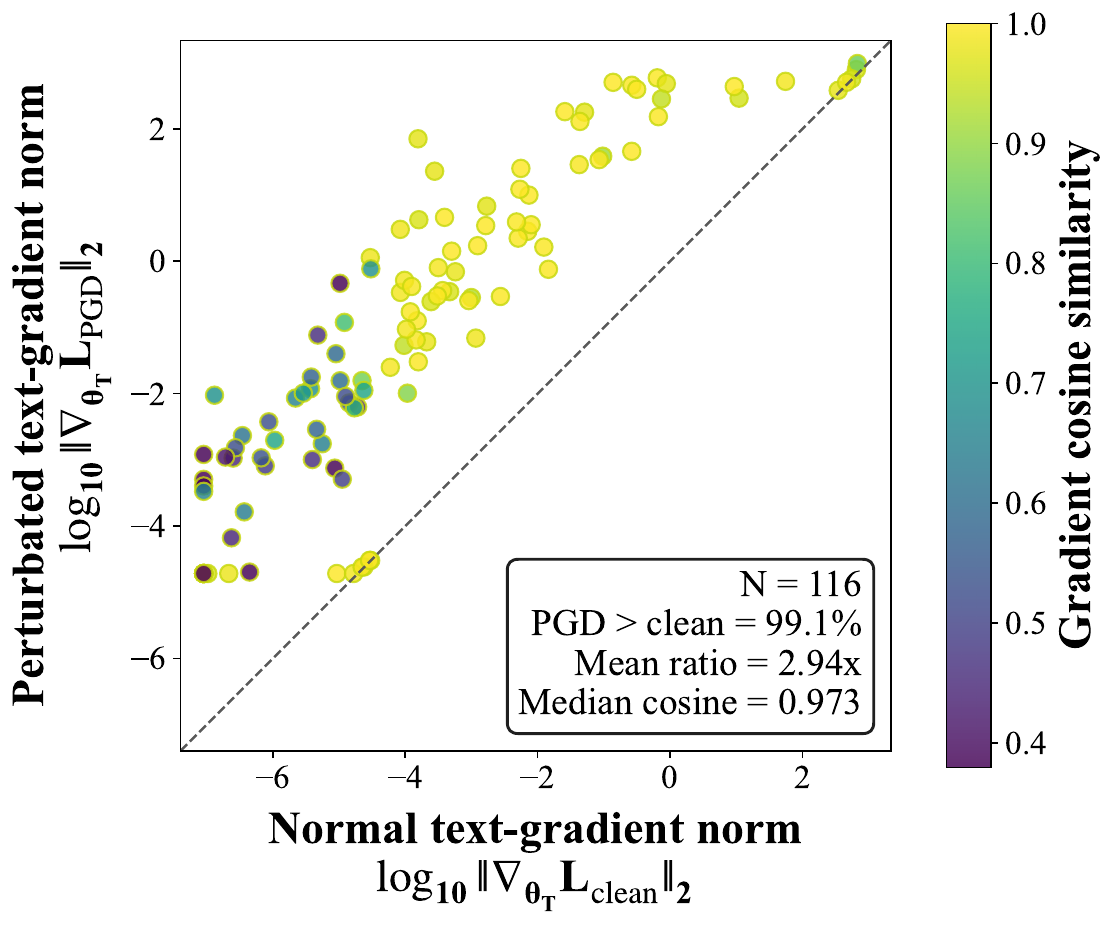}%
}
\captionof{figure}{Text-encoder gradient-flow analysis.}
\label{fig:gradient-flow}
\end{minipage}
without visual adversarial perturbations. The results are shown in Figure~\ref{fig:gradient-flow}, where each point represents a poisoned sample. Intuitively, points above the diagonal indicate that a poisoned sample produces a larger text encoder gradient when visual adversarial perturbations are applied than when they are not. We observe that 99.1\% of the poisoned samples lie above the diagonal, with gradient norms amplified by an average factor of 2.94. Meanwhile, the median cosine similarity between the gradients under the two settings is 0.973. These results provide empirical evidence that, although visual adversarial perturbations do not directly modify the textual input, they can significantly influence the model’s learning of the textual modality.

\subsection{Cross-Model Generalization}
\label{appendix-generalization}

\subsubsection{Generalization to Different CLIP Backbones.}
\label{appendix-different-clip}
\begin{table}[t]
  \centering
  \small
  \setlength{\tabcolsep}{3.2pt}
  \renewcommand{\arraystretch}{0.98}
  \begin{tabular}{@{}lcccc@{}}
    \toprule
    Backbone & R@1 & R@5 & R@10 & R@50 \\
    \midrule
    RN50x4 (Default) & 44.29 & 73.52 & 82.65 & 95.43 \\
    RN50             & 38.36 & 69.86 & 77.17 & 92.69 \\
    ViT-B/32         & 41.55 & 74.43 & 83.56 & 98.17 \\
    ViT-B/16         & 46.58 & 75.34 & 87.21 & 97.72 \\
    \bottomrule
  \end{tabular}
  \normalsize
  \caption{ASR of different CLIP backbones on CIRR.}
  \label{performance-backbone}
\end{table}

We first investigate the generalization ability of TGB across different CLIP backbones. Specifically, we conduct experiments under Attack \uppercase\expandafter{\romannumeral 2} on CIRR using four different CLIP backbones: RN50×4, RN50, ViT-B/16, and ViT-B/32. For each backbone, we vary the perturbation direction ($\lambda = \pm 1$) and the perturbation budget $\epsilon$. The experimental results are summarized in Table~\ref{performance-backbone} and Figure~\ref{fig:backbone_combined}. We observe that, without adversarial perturbations ($\epsilon = 0$), different CLIP backbones achieve comparable ASR (in Table~\ref{performance-backbone}), indicating that the proposed TGB generalizes well across backbones. In addition, the ASR curves of R@1 and R@5 under different perturbation directions and budgets, as shown in Figure~\ref{fig:backbone_combined}, exhibit consistent trends across all four backbones. This demonstrates that the effect of visual adversarial perturbations is also consistent across different CLIP backbones.

\subsubsection{Generalization to BLIP-2}
\label{appendix-blip}

\begin{table}[!htbp]
    \centering
    \small
    \setlength{\tabcolsep}{2.2pt}
    \renewcommand{\arraystretch}{0.98}
    \begin{tabular}{@{}llccccc@{}}
        \toprule
        Metric & Evaluation & Clean & I & II & III & IV \\
        \midrule
        \multirow{2}{*}{R@10}
        & Utility & 50.12 & 50.50 & 49.50 & 48.06 & 46.41 \\
        & ASR     & 0.00  & 99.39 & 22.09 & 14.69 & 1.02 \\
        \midrule
        \multirow{2}{*}{R@50}
        & Utility & 72.31 & 72.18 & 70.94 & 70.19 & 68.62 \\
        & ASR     & 0.00  & 100.00 & 66.25 & 52.58 & 2.43 \\
        \bottomrule
    \end{tabular}
    \normalsize
    \caption{Utility and ASR of
    BLIP-2-based models on FashionIQ.}
    \label{tab:blip2-results}
\end{table}

\noindent
\begin{minipage}[t]{0.42\columnwidth}%
\hspace*{\savedparindent}We further conduct experiments on FashionIQ using a BLIP-2-based CIR model (\texttt{blip2-cir-cat}) with an image–text contrastive (ITC) retrieval objective. We follow the attack setting in Appendix~\ref{appendix:default_attack_settings} and train clean models and backdoored models under Attacks \uppercase\expandafter{\romannumeral 1}–\uppercase\expandafter{\romannumeral 4}. The evaluation results are shown in Table~\ref{tab:blip2-results}. We 
\end{minipage}\hfill
\begin{minipage}[t]{0.56\columnwidth}
\centering
\raisebox{\dimexpr\ht\strutbox-\height\relax}{%
  \includegraphics[width=0.96\linewidth]{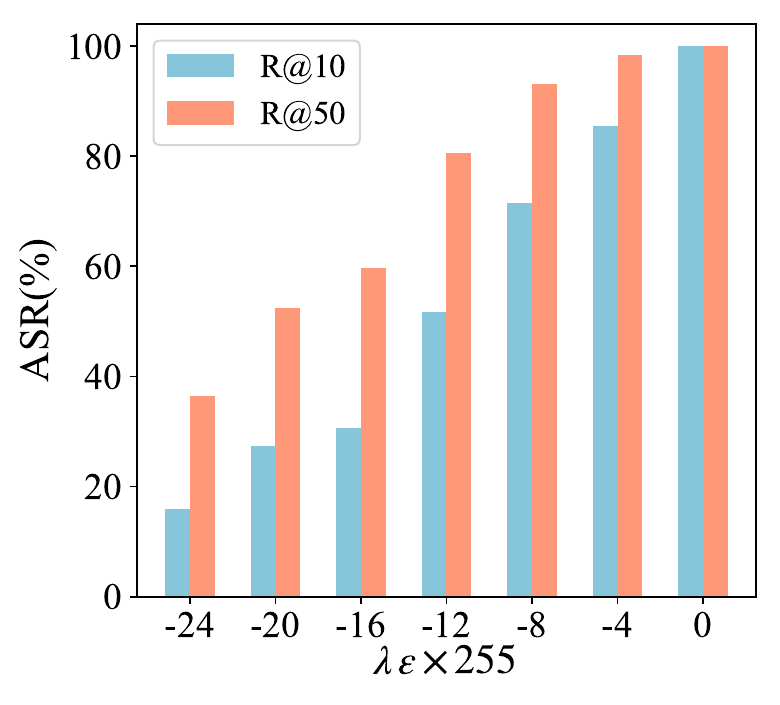}%
}
\captionof{figure}{Effect of visual adversarial perturbations on BLIP-2-based models.}
\label{fig:blip2-perturbation}
\end{minipage}

\noindent
can observe that the ASR trends of TGB and the preservation of model utility are consistent with those observed on CLIP-based models.

We then study the effect of visual adversarial perturbations on the attack strength of TGB by varying the perturbation budget under Attack \uppercase\expandafter{\romannumeral 1} ($\lambda = -1$). The results are shown in Figure~\ref{fig:blip2-perturbation}. We can observe that as the perturbation budget increases from $\epsilon = 0$ to $\epsilon = 24/255$, the R@10 ASR decreases from 99.39\% to 16.0\%, and the R@50 ASR decreases from 100.0\% to approximately 36.5\%. This trend is also consistent with that observed in CLIP-based models.

Overall, the BLIP-2-based results exhibit trends consistent with the corresponding CLIP results in terms of model utility, attack effectiveness of TGB, and the effect of visual adversarial perturbations. These results indicate that our method is not restricted to CLIP-based models and suggest its potential applicability to other multimodal pretrained models with different architectures and training objectives.

\subsection{Possible defenses}
\label{appendix:defense}
In this section, we explore possible defenses against TGB. Following \citet{yang2023data}, we discuss the effects of two types of defenses, pre-training and post-training, on the proposed backdoor attack.

\subsubsection{Pre-training Defenses}
\label{appendix:pretraining-defense}
Pre-training defenses refer to dataset-level methods that aim to filter out poisoned samples from the training data. Specifically, such defenses typically employ an auxiliary model to compute the cosine distance between clean and poisoned samples, and then apply a threshold to remove suspicious data. In our attack, the trigger is a commonly occurring natural word in the text, and the attacker-specified target output is typically semantically related to the trigger word. As a result, poisoned samples can remain largely indistinguishable from clean samples.

To evaluate the effect of pre-training defenses against TGB, we conduct experiments on FashionIQ. Specifically, we construct poisoned samples using the trigger–target pair \textit{tank2tanklike}, and then employ a pretrained CLIP model (ViT-B/16) to compute the cosine distance between clean and poisoned samples. The results are shown in Figure~\ref{fig:defense1}. It can be seen that the cosine distance distributions of clean and poisoned samples largely overlap, making it difficult to distinguish poisoned samples from clean ones via thresholding. These results demonstrate that TGB can effectively bypass pre-training defense.

\subsubsection{Post-training Defenses}
\label{appendix:post-training-defense}
Post-training defenses aim to eliminate backdoor effects by further fine-tuning the poisoned model on clean data. We evaluate the effectiveness of post-training defenses against TGB under Attack \uppercase\expandafter{\romannumeral 1} on the CIRR dataset. Specifically, we first train the model on the poisoned training set for 20 epochs, followed by an additional 5 epochs of fine-tuning on the clean training set as the post-training defense. We consider two backdoored models: one trained without visual adversarial perturbations and the other trained with visual adversarial perturbations ($\lambda = +1$ and $\epsilon = 24/255$). We report the model’s ASR measured by Recall@5, with the corresponding learning curves shown in Figure~\ref{fig:defense2}. We observe that, for the backdoored model without visual adversarial perturbations, the ASR rapidly decreases from 100\% to approximately 10\% after applying the post-training defense. In contrast, for the backdoored model with visual adversarial perturbations, the degradation of ASR is substantially mitigated, with the ASR consistently remaining above 60\%. This suggests that visual adversarial perturbations strengthen the learned association between the textual trigger and the target output, making the backdoor behavior more robust against post-training defenses.